\documentclass{aastex62}
\usepackage{amsmath,amssymb,CJK,soul}
\usepackage{longtable}
\usepackage{hyperref}
\usepackage{longtable}
\usepackage{threeparttablex} 
\usepackage{url}

\received{2019 November 5}
\revised{--}
\accepted{--}
\submitjournal{ApJ}

\shorttitle{Characterization of the Nucleus, Morphology and Activity of 2I/Borisov}
\shortauthors{Bolin et al.}

\begin{document}
\begin{CJK*}{UTF8}{gbsn}

\title{Characterization of the Nucleus, Morphology and Activity of Interstellar Comet 2I/Borisov by Optical and Near-Infrared GROWTH, Apache Point, IRTF, ZTF and Keck Observations\\}

\correspondingauthor{Bryce Bolin}
\email{bbolin@caltech.edu}

\author{Bryce T. Bolin}
\affiliation{Division of Physics, Mathematics and Astronomy, California Institute of Technology, Pasadena, CA 91125, U.S.A.}
\affiliation{IPAC, Mail Code 100-22, Caltech, 1200 E. California Blvd., Pasadena, CA 91125, USA}

\author{Carey M. Lisse}
\affiliation{Johns Hopkins University Applied Physics Laboratory, Laurel, MD 20723, U.S.A.}

\author{Mansi M. Kasliwal}
\affiliation{Division of Physics, Mathematics and Astronomy, California Institute of Technology, Pasadena, CA 91125, U.S.A.}

\author{Robert Quimby}
\affiliation{Department of Astronomy, San Diego State University, 5500 Campanile Dr, San Diego, CA 92182, U.S.A.}
\affiliation{Kavli Institute for the Physics and Mathematics of the Universe (WPI), The University of Tokyo Institutes for Advanced Study, The University of Tokyo, Kashiwa, Chiba 277-8583, Japan}

\author{Hanjie Tan}
\affiliation{Institute of Astronomy, National Central University, 32001, Taiwan}

\author{Chris M. Copperwheat}
\affiliation{Astrophysics Research Institute Liverpool John Moores University, 146 Brownlow Hill, Liverpool L3 5RF, United Kingdom}

\author{Zhong-Yi Lin}
\affiliation{Institute of Astronomy, National Central University, 32001, Taiwan}

\author{Alessandro Morbidelli}
\affiliation{Universit\'{e} C\^{o}te d'Azur, Observatoire de la C\^{o}te d'Azur, CNRS, Laboratoire Lagrange, Boulevard de l'Observatoire, CS 34229, 06304 Nice cedex 4, France}

\author{Lyu Abe}
\affiliation{Universit\'{e} C\^{o}te d'Azur, Observatoire de la C\^{o}te d'Azur, CNRS, Laboratoire Lagrange, France}

\author{Philippe Bendjoya}
\affiliation{Universit\'{e} C\^{o}te d'Azur, Observatoire de la C\^{o}te d'Azur, CNRS, Laboratoire Lagrange, France}

\author{Kevin B. Burdge}
\affiliation{Division of Physics, Mathematics and Astronomy, California Institute of Technology, Pasadena, CA 91125, U.S.A.}

\author{Michael Coughlin}
\altaffiliation{David and Ellen Lee Prize Postdoctoral Fellow}
\affiliation{LIGO Laboratory, West Bridge, Rm. 257, California Institute of Technology, MC 100-36, Pasadena, CA 91125, U.S.A.}

\author{Christoffer Fremling}
\affiliation{Division of Physics, Mathematics, and Astronomy, California Institute of Technology, Pasadena, CA 91125, U.S.A.}

\author{Ryosuke Itoh}
\affiliation{Bisei Astronomical Observatory, 1723-70 Ohkur a, Bisei, Ibara, Okayama, 714-1411, Japan}

\author{Michael Koss}
\affiliation{Eureka Scientific, 5248 Valley View Rd, El Sobrante, CA 94803, U.S.A.}

\author[0000-0002-8532-9395]{Frank J. Masci}
\affiliation{IPAC, Mail Code 100-22, Caltech, 1200 E. California Blvd., Pasadena, CA 91125, USA}

\author{Syota Maeno}
\affiliation{Bisei Astronomical Observatory, 1723-70 Ohkura, Bisei, Ibara, Okayama, 714-1411, Japan}

\author{Eric E. Mamajek}
\affiliation{Jet Propulsion Laboratory, California Institute of Technology, 4800 Oak Grove Drive, Pasadena, CA 91109, U.S.A.}
\affiliation{Department of Physics \& Astronomy, University of Rochester, Rochester, NY 14627, U.S.A.}

\author{Federico Marocco}
\affiliation{Jet Propulsion Laboratory, California Institute of Technology, 4800 Oak Grove Drive, Pasadena, CA 91109, U.S.A.}
\affiliation{IPAC, Mail Code 100-22, Caltech, 1200 E. California Blvd., Pasadena, CA 91125, USA}

\author{Katsuhiro Murata}
\affiliation{Tokyo Institute of Technology, 2 Chome-12-1 Ookayama, Meguro City, Tokyo 152-8550, Japan}

\author{Jean-Pierre Rivet}
\affiliation{Universit\'{e} C\^{o}te d'Azur, Observatoire de la C\^{o}te d'Azur, CNRS, Laboratoire Lagrange, France}

\author{Michael L. Sitko}
\affiliation{Department of Physics, University of Cincinnati, Cincinnati, OH 45221-0011} 
\affiliation{Space Science Institute, Boulder, CO 80301, U.S.A.} 

\author{Daniel Stern}
\affiliation{Jet Propulsion Laboratory, California Institute of Technology, 4800 Oak Grove Drive, Pasadena, CA 91109, U.S.A.}

\author{David Vernet}
\affiliation{Universit\'{e} C\^{o}te d'Azur, Observatoire de la C\^{o}te d'Azur,
      UMS Galil\'{e}e, France}

\author{Richard Walters}
\affiliation{Division of Physics, Mathematics, and Astronomy, California Institute of Technology, Pasadena, CA 91125, U.S.A.}

\author{Lin Yan}
\affiliation{Division of Physics, Mathematics, and Astronomy, California Institute of Technology, Pasadena, CA 91125, U.S.A.}

\author{Igor Andreoni}
\affiliation{Division of Physics, Mathematics and Astronomy, California Institute of Technology, Pasadena, CA 91125, U.S.A.}

\author{Varun Bhalerao}
\affiliation{Department of Physics, Indian Institute of Technology Bombay, Powai, Mumbai-400076, India}

\author[0000-0002-2668-7248]{Dennis Bodewits}
\affiliation{Physics Department, Leach Science Center, Auburn University, Auburn, AL 36832, U.S.A.}

\author{Kishalay De}
\affiliation{Division of Physics, Mathematics and Astronomy, California Institute of Technology, Pasadena, CA 91125, U.S.A.}

\author{Kunal P. Deshmukh}
\affiliation{Department of Metallurgical Engineering and Materials Science, Indian Institute of Technology Bombay, Powai, Mumbai-400076, India}

\author[0000-0001-8018-5348]{Eric C. Bellm}
\affiliation{DIRAC Institute, Department of Astronomy, University of Washington, 3910 15th Avenue NE, Seattle, WA 98195, USA}

\author[0000-0003-0901-1606]{Nadejda Blagorodnova}
\affiliation{Department of Astrophysics/IMAPP, Radboud University, Nijmegen, The Netherlands}

\author{Derek Buzasi}
\affiliation{Department of Chemistry and Physics, Florida Gulf Coast University, 10501 FGCU Blvd S, Fort Myers, FL 33965, U.S.A.}

\author{S. Bradley Cenko}
\affiliation{Astrophysics Science Division, NASA Goddard Space Flight Center, 8800 Greenbelt Road, Greenbelt, MD 20771, U.S.A.}
\affiliation{Joint Space-Science Institute, University of Maryland, College Park, MD 20742, U.S.A.}

\author[0000-0003-1656-4540]{Chan-Kao Chang}
\affiliation{Institute of Astronomy, National Central University, 32001, Taiwan}

\author{Drew Chojnowski}
\affiliation{Department of Astronomy,New Mexico State University, PO Box 30001, MSC 4500, Las Cruces, NM 88001, U.S.A.}

\author{Richard Dekany}
\affiliation{Caltech Optical Observatories, California Institute of Technology, Pasadena, CA 91125, U.S.A.}

\author[0000-0001-5060-8733]{Dmitry A. Duev}
\affiliation{Division of Physics, Mathematics, and Astronomy, California Institute of Technology, Pasadena, CA 91125, USA}

\author{Matthew Graham}
\affiliation{Division of Physics, Mathematics, and Astronomy, California Institute of Technology, Pasadena, CA 91125, U.S.A.}

\author{Mario Juri\'{c}}
\affiliation{Department of Astronomy, University of Washington, 3910 15th Ave NE, Seattle, WA 98195, U.S.A.}

\author[0000-0001-5390-8563]{Shrinivas R. Kulkarni}
\affiliation{Division of Physics, Mathematics, and Astronomy, California Institute of Technology, Pasadena, CA 91125, U.S.A.}

\author[0000-0002-6540-1484]{Thomas Kupfer}
\affiliation{Kavli Institute for Theoretical Physics, University of California, Santa Barbara, CA 93106, U.S.A.}

\author[0000-0003-2242-0244]{Ashish Mahabal}
\affiliation{Division of Physics, Mathematics, and Astronomy, California Institute of Technology, Pasadena, CA 91125, U.S.A.}
\affiliation{Center for Data Driven Discovery, California Institute of Technology, Pasadena, CA 91125, U.S.A.}

\author{James D. Neill}
\affiliation{Division of Physics, Mathematics, and Astronomy, California Institute of Technology, Pasadena, CA 91125, U.S.A.}

\author{Chow-Choong Ngeow}
\affiliation{Institute of Astronomy, National Central University, 32001, Taiwan}

\author{Bryan Penprase}
\affiliation{Soka University of America, 425 Pauling Hall, 1 University Drive, Aliso Viejo, CA 92656, U.S.A.}

\author{Reed Riddle}
\affiliation{Caltech Optical Observatories, California Institute of Technology, Pasadena, CA 91125}

\author{Hector Rodriguez}
\affiliation{Caltech Optical Observatories, California Institute of Technology, Pasadena, CA 91125, U.S.A}

\author{Roger M. Smith}
\affiliation{Caltech Optical Observatories, California Institute of Technology, Pasadena, CA 91125}

\author{Philippe Rosnet}
\affiliation{Universit\'{e} Clermont Auvergne, CNRS/IN2P3, LPC, Clermont-Ferrand, France}

\author{Jesper Sollerman}
\affiliation{Department of Astronomy, Stockholm University, SE-106 91 Stockholm, Sweden}

\author[0000-0001-6753-1488]{Maayane T. Soumagnac}
\affiliation{Lawrence Berkeley National Laboratory, 1 Cyclotron Road, Berkeley, CA 94720, U.S.A.}
\affiliation{Department of Particle Physics and Astrophysics, Weizmann Institute of Science, Rehovot 76100, Israel}





\begin{abstract}
We present visible and near-infrared photometric and spectroscopic observations of interstellar object 2I/Borisov taken from 2019 September 10 to 2019 December 20 using the GROWTH, the APO ARC 3.5 m and the NASA/IRTF 3.0 m combined with post and pre-discovery observations of 2I obtained by ZTF from 2019 March 17 to 2019 May 5. Comparison with imaging of distant Solar System comets shows an object very similar to mildly active Solar System comets with an out-gassing rate of $\sim$10$^{27}$ mol/sec. The photometry, taken in filters spanning the visible and NIR range shows a gradual brightening trend of $\sim0.03$ mags/day since 2019 September 10 UTC for a reddish object becoming neutral in the NIR. The lightcurve from recent and pre-discovery data reveals a brightness trend suggesting the recent onset of significant H$_2$O sublimation with the comet being active with super volatiles such as CO at heliocentric distances $>$6 au consistent with its extended morphology. Using the advanced capability to significantly reduce the scattered light from the coma enabled by high-resolution NIR images from Keck adaptive optics taken on 2019 October 04, we estimate a diameter of 2I's nucleus of $\lesssim$1.4 km. We use the size estimates of 1I/'Oumuamua and 2I/Borisov to roughly estimate the slope of the ISO size-distribution resulting in a slope of $\sim$3.4$\pm$1.2, similar to Solar System comets and bodies produced from collisional equilibrium.
\end{abstract}

\section{Introduction}
The study of Interstellar Objects (ISOs) is presently the best opportunity to directly observe the contents of extra-solar circumstellar disks at larger than cm-size scales. Present-day observations are limited to observing the micron-sized \citep[e.g.,][]{Lisse2012, Lisse2017} to millimeter-sized \citep[][]{MacGregor2019} dust contents of extra-solar disks. Indirect observations of macroscopic objects and their volatile contents in debris disks can be obtained through the massive amounts of dust produced by their collision with each other \citep[][]{Meng2014,Su2019}, their presence around young stars \citep[][]{Chen2006} or sometimes by their transit of stars \citep[][]{Rappaport2018}, but observing and obtaining the physical properties and volatile contents of specific bodies from other stars has remained elusive.

2I/Borisov (2I) is the second example of a macroscopic body with a definitive interstellar origin to be discovered, discovered on August 30th, 2019 by amateur astronomer Gennadiy Borisov, the hyperbolic orbit with $e\simeq$ 3.35 was confirmed on September 11, 2019 \citep[][]{Williams2019a}. Unlike the first interstellar object to be discovered, 1I/'Oumuamua, \citep[][]{Williams2017} which did not have a cometary appearance in ground-based \citep[][]{Bolin2018,Jewitt2017a} or space-based images \citep[][]{Micheli2018}, Borisov has a distinct comet-like appearance with a diffuse coma \citep[][]{Jewitt2019}. This provides an opportunity to characterize the properties of a cometary interstellar body for the first time.

Initial spectroscopic observations have revealed the presence of CN and C$_2$ gas in the coma of 2I with gas production rates comparable to Solar System comets at similar, heliocentric distances, r{$_h$} \citep[][]{Fitzsimmons2019, Kareta2019,Opitom2019}. Using Solar System comets as a guide, the production rate of CN observed in 2I implies a nuclear diameter of $\sim$6 km. The measured size combined with canonical models describing the brightness of 2I driven by H$_2$O or CO sublimation produces very different results versus heliocentric distance, as a body dominated by CO sublimation will be active much farther away from the Sun due to CO's much lower enthalpy of sublimation \citep[][]{Meech2004, Fitzsimmons2019}. Therefore, it may be possible to distinguish between different compositional models of 2I by measuring its brightness at different heliocentric distances covering a wide span of times \citep[e.g.,][]{Meech2017b,Jewitt2017cc}. This indeed appears to be the case, with ZTF precovery observations of 2I strongly favoring the activity of the comet being driven by more volatile species other than H$_2$0, such as CO or CO$_2$ \citep{Ye2019b}.
In this paper, we build upon these ZTF results and present visible and near-infrared observations of 2I, its morphology, the null result for variability on short term timescales, estimates of the comet's size, af$\rho$ and dust mass-loss rate, strengthened evidence for activity driven by CO and H$_2$O and an estimate of the ISO cumulative size distribution slope.

\section{Observations}

Since before the official announcement of the hyperbolic orbit of 2I, optical observations were being taken to characterize the object's brightness and refine its orbit. We used the rapid-response capability of the GROWTH (Global Relay of Observatories Watching Transients Happen) network to organize and schedule observations of 2I. Observations were done at different observatories around the world, all conducted at high airmass, $\gtrsim$2, just before or during astronomical twilight owing to the small, $43^{\circ}$ solar elongation of the comet in mid-September 2019. In addition to the difficulty of observing near twilight and at high airmass, the comet had a fast sky motion of $\sim$1\arcsec /min, necessitating the use of non-sidereal tracking for the majority of the observations. \\

We present here the observations of a monitoring campaign lead through the GROWTH collaboration \citep[][]{Kasliwal2019} combined with data from Apache Point Observatory's Astrophysical Research Consortium (ARC) 3.5 m telescope, the NASA/IRTF'S 3.0 m telescope, and Zwicky Transient Facility (ZTF) and Keck Observatory. The time span of our observations is between 2019 March 17 and 2019 December 20 UTC.

\subsection{SED Machine}
The first observations of 2I used in this study were made with the SED Machine (SEDM), operating on the P60 telescope on Palomar \citep[][]{Blagorodnova2018, Rigault2019}. The SEDM possesses a multi-band CCD camera that we used to obtain SDSS $r$-band images in 60 s exposures of 2I on 2019 September 10 and 2019 September 11 UTC. The telescope was tracked non-sidereally according to the sky motion of 2I resulting in background stars that were trailed $\sim$2\arcsec. The astrometric positions of 2I were computed and submitted to the MPC to refine the object's orbit \citep[][]{Williams2019b}. The airmass at the time of the observations was $\sim$2 and the seeing was $\sim$1.4\arcsec~in the images taken for the object. This facility is a member of the GROWTH collaboration.

\subsection{Apache Point Astrophysical Research Consortium 3.5 m}
Immediately following the MPC's announcement of the discovery of 2I, we obtained director's discretionary time to observe 2I with the Apache Point Observatory's ARC 3.5 m. The first observations with the ARC 3.5 m were made on 2019 September 12 UTC in photometric conditions with the ARCTIC large-format optical CCD camera \citep[][]{Huehnerhoff2016}. The camera was used in full-frame, quad amplifier readout, 2$\times$2 binning mode resulting in a pixel scale of 0.228\arcsec. Exposures were each 120 s long made in a rotating order of four filters, SDSS $griz$ in order to mitigate the potential effects of rotational variability on the color calculations \citep[e.g.,][]{Hanus2018}, and were dithered by 20\arcsec between exposures of the same filter. In total, five $g$, eight $r$, one $i$ and two $z$ exposures were obtained. The telescope was tracked at the sky-motion rate of the comet resulting in stars that were trailed by $\sim$2\arcsec. Additional observations were made on 2019 September 27 UTC using the Aspen Apogee Camera in $R$-band and on 2019 October 12 UTC using the ARCTIC camera with $Bgriz$ filters. Seeing was exceptionally good, $\sim$0.55\arcsec~ in the images taken for the object, on the night of the 2019 September 12 UTC, however, the observations were conducted at high airmass and into astronomical twilight reducing the sensitivity of the observations.

The ARC 3.5 m was also used to obtain SDSS/Maunakea $zJHK$ photometry of 2I on 2019 September 19 and 2019 September 27 UTC with the NIC-FPS near-infrared (NIR) camera \citep[][]{Vincent2003}. A revolving $zJHK$ filter sequence was used with a five-point dither pattern. To avoid the effects of the high sky background in the NIR, 40 s and 20 s exposures were used for the $H$ and $K$ filter images, respectively, and 120 s and 60 s exposures were used for the $z$ and $J$ filter images. Up to eight Fowler samples were used per readout to limit read-out noise. Seeing was $\sim$1\arcsec~or better during the nights of 2019 September 12, 2019, September 19, 2019, September 27, 2019, October 12 and 2019 October 21 UTC.

\subsection{Lulin Optical Telescope}
Also soon after the discovery of 2I, imaging data were obtained on 2019 September 12 UTC with the 1 m Lulin Optical Telescope (LOT) using the 2K $\times$ 2K SOPHIA camera \citep[][]{Kinoshita2005} at Lulin Observatory. Data were taken in Johnson-Cousins $V$, $B$, $R$ and $I$ bands, and the telescope was tracked non-sidereally at the comet's sky motion rate. The seeing during the observations was $\sim$3.5\arcsec~ in the images taken for the object and the airmass was $\sim$2.36.

\subsection{Bisei Observatory 101 cm}
Images of 2I were obtained at Bisei Observatory\footnote{\url{http://www.bao.city.ibara.okayama.jp/eng/sisetu.htm}} on 2019 September 15 UTC using the 101 cm reflecting telescope. Images with 60 s exposure in Johnson-Cousins $R$-band were obtained using the Astrocam optical camera, and the telescope was tracked at a sidereal rate. Seeing at the time of observations was typically $\sim$2\arcsec~in the images taken for the object and the airmass was $\sim$2. This facility is a member of the GROWTH collaboration.

\subsection{Liverpool Telescope}
On eight separate nights between 2019 September 18 and 2019 October 15 UTC, observations of 2I were obtained with the 2 m Liverpool Telescope located at the Observatorio del Roque de los Muchachos. Images were obtained using the IO:O wide-field camera with a 2x2 binning and the SDSS $g$ and $r$ filters \citep[][]{Steele2004}. A 30 second exposure time was used with the telescope tracking the target in a non-sidereal mode. Debiasing and flat fielding of the data was performed using the automated IO:O pipeline software. Seeing was typically $\sim$1\arcsec~in the images taken for the object during the observations and the airmass was $\sim$1.8-2.0. This facility is a member of the GROWTH collaboration.

\subsection{Mount Laguna Observatory 40-inch Telescope}
Optical images were obtained with the 1.0 m Telescope at the Mount Laguna Observatory \citep[][]{Smith_Nelson1969}
on 2019 September 19, 2019 September 30, 2019 October 04, 2019 October 08, 2019 October 12 and 2019 October 17 UTC. 
The E2V 42-40 CCD Camera was used to obtain typically six 90 s exposures in each of the Johnson-Cousin $V$ and $R$ filters each night.
Both sidereal and non-sidereal tracking was used, and these produced similar results due to the shortness of the exposures. The seeing during observations was typically $\lesssim$3\arcsec~as measured using stars in the images and the airmass was $\sim$1.5-2.0. This facility is a member of the GROWTH collaboration.

\subsection{NASA/Infrared Telescope Facility}
On 2019 September 20, 2019, September 22, 2019, September 29, and 2019 October 02 UTC, observations of 2I were obtained with the 3 m NASA/Infrared Telescope Facility (IRTF) Telescope located at Maunakea, Hawaii. $H$ filter images and NIR spectra were obtained with a wide 0.8\arcsec~slit and the low-resolution prism mode of the SpeX prism instrument \citep[][]{Rayner2003}, and $r$ band imagery was also obtained with the MORIS guider camera. The telescope was tracked at comet rate, and long 120 sec exposures were used. The slit was rotated to the optimal azimuth angle in order to reduce differential atmospheric refraction, which would affect the shortest wavelengths in the spectrum the most, when taking prism data for Borisov and the standard star. Because of the comet's brief time above the horizon, only 4 ABBA pairs were obtained on a typically good weather night; the evening of 2019 September 22 UTC was clouded out and little useful data obtained. The most useful data came from the observations on 2019 September 29 UTC which are described below. The airmass during time of observation was $\sim$1.5-2.0. The seeing was $\sim$0.8\arcsec~measured at zenith and was worse by $\sim$30$\%$ at the airmass of our observations.

\subsection{Zwicky Transient Facility}
ZTF is a wide field all-sky survey using Palomar Observatory's P48 Oschin Schmidt telescope \citep[][]{Bellm2019}. The mission of the ZTF survey is to discover transients which include asteroids and comets \citep[][]{Graham2019}. The ZTF camera has a 47 sq. deg field of view and can reach $r\sim$20.5 to a SNR = 5 depth in a 30 s exposure, enabling the survey to cover 3,800 sq. deg./ h. In addition to the GROWTH and APO data, we use pre-discovery observations found in the Zwicky Transient Facility (ZTF) database to extend the time range of our observations \citep[][]{Masci2019}. Seeing was typically $\sim$2\arcsec~ and the airmass was $\sim$2.

Using the latest orbital solution for 2I that was available on 2019 October 02 UTC \citep[][]{Williams2019bb}, we used the ZTF database search tool \citep[][]{Masci2019} to locate images that had overlapping coverage with the trajectory of 2I. The positional uncertainty of 2I in images as far back as March and May 2019 was less than few 10\arcsec-30\arcsec. With such a small search area, it became viable to visually spot the detections of the comet in the images where automated software would have missed these detections i.e., for being too faint, SNR $\simeq$2-3. Therefore, we searched for the detections of 2I by eye in each set of images between 2019 March 17 and 2019 May 05 UTC using the nominal position from JPL HORIZONS as a starting point. The individual detections were very weak, of the order of SNR$~$2-3 and in a high sky background owing to the fact that some of them came from observations taken during astronomical twilight.

We identified the pre-discoveries on the dates 2019 March 17, 2019, March 18, 2019, May 02 and 2019 May 05 UTC during the public and partnership surveys \citep[][]{Bellm2019b}. We used images that were taken with a 30 s $r$ filter exposure for the pre-discovery images. We used two exposures taken on 2019 March 17 UTC, two exposures taken on 2019 March 18 UTC, six exposures taken on 2019 May 02 UTC and four exposures taken on 2019 May 05 UTC \citep[][]{Ye2019b}. In addition to the pre-discovery detections, we identified additional post-discovery detections of 2I in ZTF survey data between 2019 September 11 and 2019 December 20 UTC that were taken in SDSS-like $g$ and $r$ filters \citep[][]{Graham2019}. Seeing was typically $\sim$2\arcsec~in the images taken for the object and the airmass ranged between 1.2 and 2 in the pre-discovery images.

\subsection{Keck I Telescope}
We obtained high-resolution images of 2I on 2019 October 04 UTC with the Keck I instrument OSIRIS in imaging mode using laser guide star adaptive optics (AO) \citep[][]{Larkin2006}, the first time this instrument and telescope combination had been used to track and observe a comet.  The comet was at a heliocentric distance of 2.48 au, a topo-centric distance of 2.96 au and a phase angle of 18.65$^{\circ}$ during our observations. Four 60 s exposures were made in $K_p$ band using the laser guide system with a $r\sim$15 mag star within 60\arcsec~of the comet during the observations. Because of Keck I's 33$^{\circ}$ elevation constraint in the azimuth range of 2I, observations had to wait until astronomical twilight to begin. The $K_p$ filter\footnote{\url{https://www2.keck.hawaii.edu/inst/osiris/technical/filters/filter_index.html}} is a near-infrared filter similar to the 2MASS $K_s$ filter with a central wavelength of 2114.45 nm and a FWHM bandpass of 307.03 nm. A nearby $r\sim$15 star was used with the laser guidance system for adaptive optics correction while tracking at the sky motion rate of 2I, however, the high airmass of the observations and performance of the laser system resulted in lower image quality than usual. The PSF FWHM of background stars in OSIRIS images is 0.22-0.26\arcsec~measuring in the perpendicular direction of the direction of motion of 2I. The airmass during the time of the observations was $\sim$1.6.

\subsection{C2PU facility 1.04 m Omicron telescope}
Observations of 2I were obtained with the Observatoire de la C\^{o}te d'Azur's C2PU 1.04 m telescope located at Calern on 2019 November 29 UTC. Images were obtained using the 4096 $\times$ 4096 SBIG STX-16803 CCD camera with 0.6\arcsec~pixel scale in $R$ bands. A 30-s exposure time was used with the telescope tracking the target in a non-sideral mode and the airmass of the observation were $\sim$2.4. The atmospheric seeing was $\sim$1.88\arcsec~in the images taken for the object. Debiasing and flat fielding was performed on the data using automated software.

\section{Results}
\label{sec:results}
\subsection{Optical Photometry and Colors}
\label{sec:photometry}
Data collected with the GROWTH and ARC 3.5 m telescopes were processed using flattened and dark-subtracted images produced by basic methods. Photometric measurements are obtained by using a circular aperture with a projected radius of 10,000 km at the topo-centric distance of the comet, typically, $\sim$5\arcsec. The typical seeing at our observing locations was typically well under $\sim$5\arcsec~measured in the images as described in Table~\ref{tab:phot}. The brightness of the comet was calibrated using the PanSTARRS catalog \cite[][]{Tonry2012,Flewelling2016}. Johnson-Cousins photometry was calibrated using the PanSTARRS catalog and the filter transformations described in \citet[][]{Tonry2012}.  We calibrated the photometry with in frame stars thus accounting for varying conditions at the high airmass of our observations. Sky subtraction was done using annuli with an inner radius exceeding the extent of the coma by $\gtrsim$ 10\arcsec.

As of writing, our team has been regularly monitoring the comet's brightness since 2019 September 10 UTC to 2019 December 20 UTC with telescopes in the GROWTH network at observatories from around the world as described above. The photometric observations cover a span of wavelengths from $V$ band to $I$ and $z$. To put the photometric measurements on the same scale for comparison, individual Johnson-Cousins filters were converted to the SDSS magnitude system using the colors measured here in $griz$ and $VBRI$ filters \citep{Jordi2006}. The resulting magnitudes are listed in Table~\ref{tab:phot}.

\begin{figure}
\centering
\includegraphics[scale=0.52]{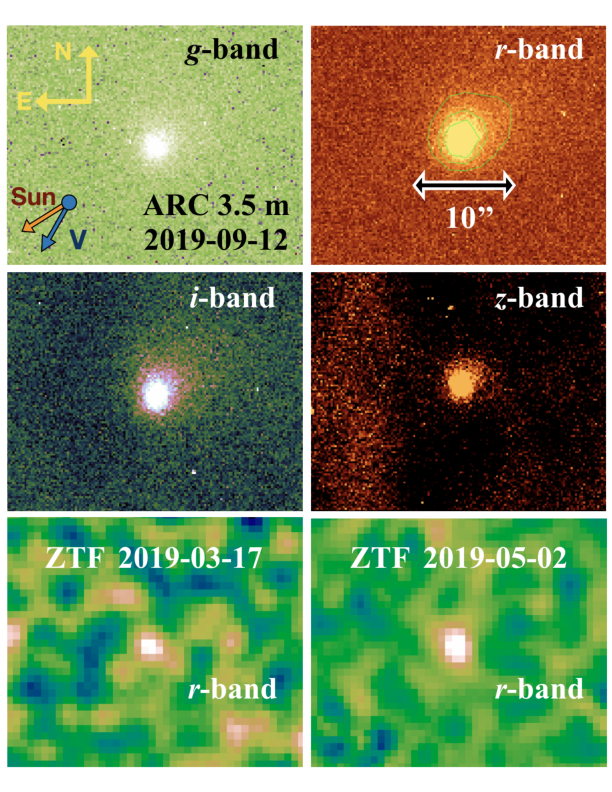}
\caption{Mosaic of $g$, $r$, $i$ and $z$ images of 2I taken with the ARC 3.5 m on 2019 September 12 UTC and pre-discovery ZTF images from 2019 March 17 and 2019 May 02 UTC. Top left panel: a composite stack of five 120 s $g$ filter exposures with the orbital velocity and solar directions. Top right panel: a composite stack of eight 120 s $r$ filter exposures and shows the extent of the comet's tail limited by sky background. Middle left panel: single 120 s $i$ filter exposure. Middle right panel: a stack of two $z$ filter exposures. The nearby background is irregular due to incomplete removal of fringes. Calibration stars were carefully chosen that were not affected by these fringe removal artifacts. Bottom left and bottom right panels: pre-discovery detections of 2I from 2019 March 17 and 2019 May 02 UTC. The 2019 March 17 UTC data is a stack of two images with an equivalent exposure time of 60 s. The 2019 May 02 UTC data is a stack of six images with an equivalent exposure time of 180 s. Both of these ZTF image stacks have been spatially smoothed to enhance faint features in the image. The artifact at the bottom of the 2019 March 17 UTC image is a star subtraction artifact. No extended coma or tail features are evident in the pre-discovery image stacks owing to the low surface brightness of these features at the time of observation. A green color scale was chosen for the ZTF $r$ filter observations to more clearly highlight these faint detections compared to the surrounding background.}
\label{fig.mosaic}
\end{figure}

A mosaic of composite images showing the detections of 2I taken by the ARC 3.5 m on 2019 September 12 UTC in $g$, $r$, $i$ and $z$ filters is shown in Fig.~\ref{fig.mosaic}. The comet has a clearly extended appearance with a diffuse tail $\sim$6.7\arcsec~long pointing in the $\sim$315$^{\circ}$ position angle. Using the stacked images taken by the ARC 3.5 m on 2019 September 12 UTC, The SDSS $griz$ filter colors of 2I are $g-r$ = 0.54 $\pm$ 0.06, $r-i$ = 0.20 $\pm$ 0.04, $i-z$ = $-$0.23 $\pm$ 0.04. Immediately after the ARC 3.5 m observations on 2019 September 12 UTC, $BVRI$ observations were obtained with the LOT resulting in the following colors: $B-V$ =0.76  $\pm$ 0.12, $V-R$ = 0.55 $\pm$ 0.09, $R-I$ = 0.37 $\pm$ 0.08. Converting the ARC 3.5 m $griz$ colors for 2I to $BVRI$ colors using the transformations in \citep[][]{Jordi2006} results in: $B-V$ = 0.69 $\pm$ 0.09, $V-R$ = 0.40 $\pm$ 0.1 and $R-I$ = 0.41 $\pm$ 0.07, which are in good agreement with the LOT $BVRI$ colors and the $BVRI$ colors obtained by \citet[][]{Fitzsimmons2019} and \citet[][]{Jewitt2019}. An additional observing run on the ARC 3.5 m was conducted on 2019 October 12 UTC where $Bgriz$ filtered observations were obtained of 2I with similar colors as measured from data obtained with the ARC 3.5 m on 2019 September 12 UTC: $g-r$ = 0.63 $\pm$ 0.05, $r-i$ = 0.20 $\pm$ 0.05 and $i-z$ = $-$0.23 $\pm$ 0.02. In addition, we calculate $B-V$ colors for data taken on 2019 October 12 UTC by converting our $g$ and $r$ measurements to a $V$ magnitude using the filter transformations for converting SDSS to Johnson-Cousins magnitudes. This results in $B-V$ = 0.68 $\pm$ 0.04, similar to the $B-V$ colors obtained by the LOT on 2019 September 12 UTC and by \citet[][]{Jewitt2019}.

\begin{figure}
\centering
\includegraphics[scale=0.52]{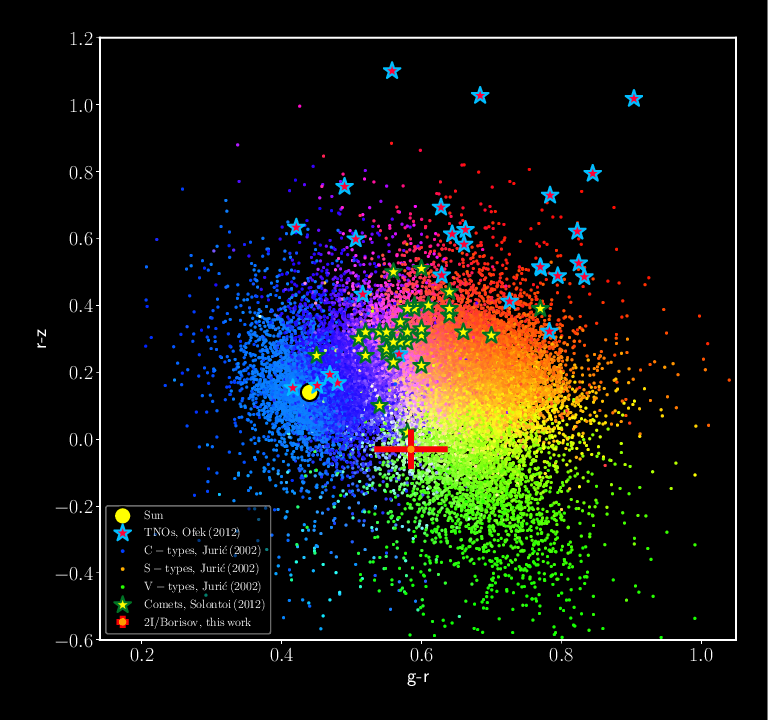}
\caption{$g-r$ vs. $r-z$ colors of 2I plotted with $g-r$ and $r-z$ colors of other Solar System bodies such as inner Solar System asteroids such as C, S and V types, \citep{Ivezic2001,Juric2002,DeMeo2013}, active comets \citep[][]{Solontoi2012} and KBOs \citep[][]{Ofek2012}. The colorization scheme of data points for asteroids by their $griz$ colors is adapted from \citet[][]{Ivezic2002}. We note that the comparison of the colors of 2I to active comets in \citep[][]{Solontoi2012} is the most appropriate comparison rather than inactive bodies since the colors of 2I are most representative of its dust rather than bare nucleus. The colors of inactive bodies are present for comparison only.}
\label{fig.grz}
\end{figure}

We extend our color analysis redward of the SDSS $i$ and Cousins $I$ filters, centered at 762 nm and 880 nm respectively, to 913 nm with the inclusion of the SDSS $z$ filter. While the visual spectrum reported by \citet[][]{deLeon2020,Hui2020a} shows an overall red appearance, our $g-r$ vs. $r-z$ colors of 2I show similarity with neutral and bluish Solar System bodies, and 2I does not appear to be as red as outer Solar System bodies such as comets and Kuiper Belt objects (KBOs) with the inclusion of the longer wavelength $z$ filter data as seen in Fig~\ref{fig.grz}. This is in contrast with the apparently slightly red color of 2I in $B-V$ vs. $V-R$ color space as seen in Fig.~5 of \citet[][]{Jewitt2019} which only goes as red as 635 nm for the $R$ vs 913 nm for the $z$ filter $g-r$ vs. $r-z$ color space. We must caution that the comparison of the colors between 2I and known solar system comets can be affected by the fact that comet dust for active comets can modify their apparent color compared to inactive bodies \citep[][]{Li2013}. We also further caution that although 2I appears neutral to reddish with the addition of longer wavelengths in the $g-r$ vs. $r-z$ color space compared to $B$-$V$ vs. $V$-$R$ color space, the interpretation of the colors of small bodies is limited by the fact that many Solar System bodies appear neutral in optical colors spanning wavelengths 477 nm to 913 nm for filters $g$ to $z$ \citep[][]{Bus2002}. However, Solar System objects that appear to be neutral in optical wavelengths can be revealed to be much more red with the inclusion of even longer wavelength data in NIR range \citep[e.g.,][]{DeMeo2009,Schwamb2019}, as further discussed in Section~\ref{s.nir}. We wish to reiterate that this comparison of colors with inactive bodies is for reference only. In addition 2I exhibits colors that are markedly different from active solar system objects, being less red in $r-z$ color that may indicate that the color of the dust of 2I is different from Solar System comets.

We increased the range of our long term lightcurve by using pre-discovery observations of 2I found in $r$ filter images from the ZTF survey spanning from March to May 2019. \cite{Ye2019b} present these measurements and analysis; here, we repeat their extraction from raw data both as a cross-check and for methodological consistency reasons. 

The pre-discovery detections from ZTF were stacked in the individual images increasing their SNR by locating detections in several overlapping images taken on the same night that were processed to remove static sources. The limiting magnitude in the image stacks was $r\sim$21.5 for the 60 s equivalent exposure time image stacks taken in 2019 March and $r\sim$ 22.5 and 22 for the 180 s and 120 s equivalent exposure time image stacks were taken on 2019 May 2 and 2019 May 5, respectively. The image stacks showing the individual detections taken on 2019 March 17 UTC and 2019 May 2 UTC are shown in the bottom panels on Fig.~\ref{fig.mosaic}. In addition to the photometry, the pre-discovery detections were measured astrometrically and submitted to the MPC allowing for the orbital arc to be significantly extended by several months improving its accuracy, and for use by the community to study 2I. The photometry from the post-discovery observations by the GROWTH and ZTF telescopes are listed in Table~\ref{tab:phot}.

\subsection{NIR Photometry and Spectrum}
\label{s.nir}
\begin{figure}
\centering
\includegraphics[scale=0.52]{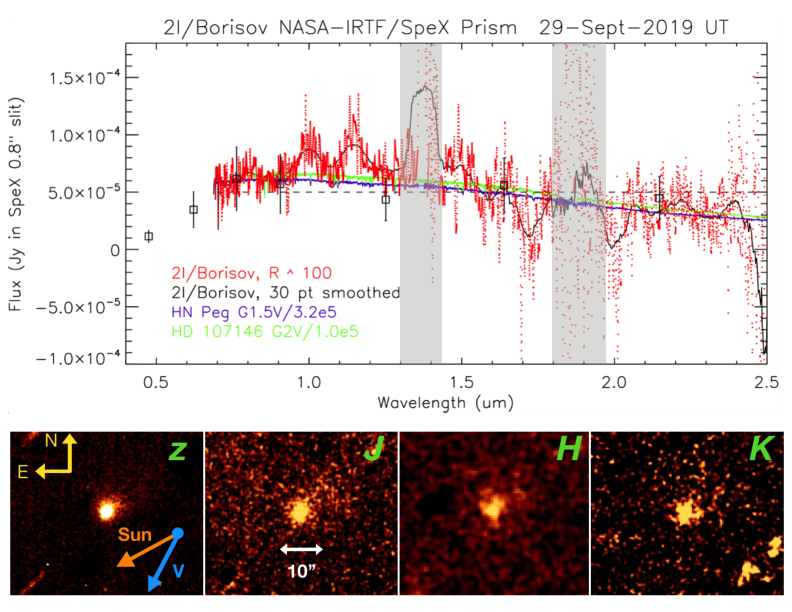}
\caption{Top panel: scaled flux IRTF SpeX spectrum of 2I taken on 2019 September 29 UTC. The red line is a $R\sim$100 spectrum between 0.7 and 2.5 microns from two A-B B-A 120 s pair subtractions. The black line is the smoothed spectrum of 2I with a 30 pt ($\sim$50 nm) running mean. The blue and green lines correspond to G1.5V and G2V analog stars HN Peg and HD 107146. The spectral energy distribution (SED) is overall reddish-neutral with some slight deviations in the 0.9-1.2 micron range. The $gri$ fluxes obtained in observations on 2019 October 12 and the $zJHK$ fluxes obtained in observations on 2019 September 27 with the ARC 3.5 m are over plotted on the spectrum and are in rough agreement with the spectrum. Emission features at $\sim$1.4 and $\sim$1.8 microns are of terrestrial atmospheric origin. Bottom panel: $zJHK$ image stacks of 2I taken on 2019 September 27 UTC with NIC-FPS on the ARC 3.5 m. The $z$ and $J$ images are a 600 s robust mean stack, the $H$ and $K$ images are 200 s robust mean stacks. All images have been spatially smoothed to enhance faint features. The north and south direction and the solar and orbital velocity directions are indicated on the $z$ band panel. Regions of the spectrum degraded by sky absorption are greyed-out.}
\label{fig.NIR}
\end{figure}

From our VisNIR observations with the ARC 3.5 cometary morphology is evident in the $z$ image, but the cometary appearance is suppressed in the longer wavelength $JHK$ images as seen in Fig.~\ref{fig.NIR} due to light scattering by cometary dust being less efficient at longer wavelengths \citep[][]{Fernandez2013,Bauer2017}. The $z$ and $JHK$ photometry were calibrated using the PanSTARRS \citep[][]{Chambers2016} and 2MASS \citep[][]{Skrutskie2006} catalogs. We measure magnitudes $z$ = 17.57 $\pm$ 0.05, $J$ = 16.80 $\pm$ 0.05, $H$ = 16.01 $\pm$ 0.09 and $K$ = 15.81 $\pm$ 0.10. Combined with the $R$ filter observation also taken on 2019 September 27 UTC, the resulting colors are, after converting the $R$ measurement to $r$ = 17.60 $\pm$ 0.04, $r-z$ = 0.03 $\pm$ 0.06, $r-J$ = 0.80 $\pm$ 0.06, $z-J$ = 0.77 $\pm$ 0.07, $J-H$ = 0.79 $\pm$ 0.10, $H-K$ = 0.20 $\pm$ 0.13 similar to neutral Solar System objects and distinct from very red outer Solar System objects \citep[][]{Schwamb2019}. As seen in Fig.~3 from \citet[][]{Bannister2017}, the rough dividing line in $r$-$J$ separating outer Solar System objects from inner Solar System objects is $r$-$J$ $\gtrsim$ 1.2 where the $r$-$J$ of 2I is $\sim$0.8.

We made a pair-subtracted stack of the four 120 s ABBA sequence SpeX prism exposures of 2I resulting in the composite spectrum seen in the top panel of Fig.~\ref{fig.NIR}. The full compliment of $gri$ and $zJHK$ photometry from 2019 October 12 and 2019 September 27 UTC are over-plotted on top of the NIR spectrum showing agreement with the visible portion with the visible spectrum of \citep[][]{deLeon2019}. The spectrum was adjusted to the photometric points. The infrared color of 2I, as determined by the continuum slope of the prism spectra, was found to be neutral-grey in agreement with the $rzJHK$ colors in contrast with \citet[][]{Yang2019}. No definitive absorption or emission lines were found within the errors of the measurements similar to the lack of emission lines seen in the spectra of Solar System comets and asteroids in the 0.7 -2.5 micron range \citep[][]{Feldman2004}. 

The colors are typical of optically reddish objects containing refractory organics and silicates that become NIR-neutral because of the presence of water ice \citep[][]{Yang2009,Snodgrass2017}. Because the flux of 2I is dominated by its coma as discussed in Section~\ref{sec:photometry}, we can infer, by analogy with solar system comet spectra, that the Borisov NIR spectrum that its coma dust contains silicates, refractory organics, and water ice \citep[][]{Protopapa2014,BockeleeMorvan2017}, though recent observations suggest that the main driver of the activity is CO and is likely responsible for driving the dust production \citep[][]{Bodewits2020}.

\begin{figure*}
\centering
\includegraphics[width=0.45\textwidth]{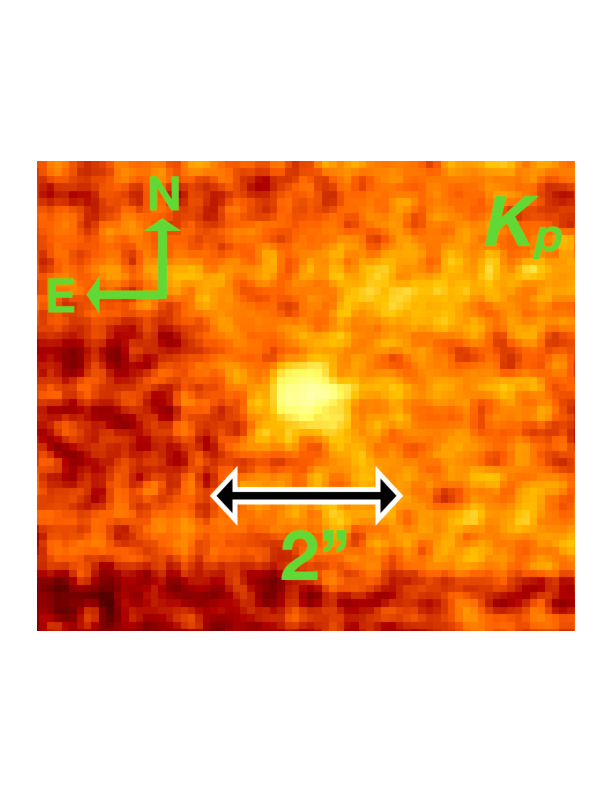}
\caption{$K_p$ image of 2I taken with the OSIRIS adaptive optics instrument on Keck I tracking at the sky motion rate of the comet. The image is a composite stack of four 60 s $K_p$ exposures stacked on the position of the comet in each individual exposure. The image has been 4$\times$4 binned giving it a pixel scale of 0.08\arcsec. The FWHM of the background stars measured perpendicularly to the rate of motion of 2I is $\sim$0.2\arcsec. The detection is PSF-like without an extended appearance or a tail visible in the image. The image has been Gaussian-smoothed by two binned pixels.}
\label{fFiig:osirisKp}
\end{figure*}

From our OSIRIS observations with Keck I, A composite stack of the $K_p$ images of 2I is shown in Fig.~\ref{fFiig:osirisKp}. No extended coma or tail features are evident in the OSIRIS images taken on 2019 October 04 UTC owing to the low surface brightness of these features in $K$ band wavelengths similar to the NIR wavelength images taken by the ARC 3.5 m on 2019 September 27 UTC as seen in the bottom panel of Fig.~\ref{fig.NIR}. We estimate the apparent brightness of 2I in the AO $K_p$ images measured to be $m_{K_p}$ = 15.68 $\pm$ 0.06 using a 4.7\arcsec~circular aperture with a projected radius of 10,000 km at the topo-centric distance of the comet of 2.96 au on 2019 October 04 and the  zero-point of 27.6 determined for the $K_p$ of the OSIRIS instrument \footnote{\url{https://www2.keck.hawaii.edu/inst/osiris/OSIRIS_Manual_v2.2.pdf}}.

\subsection{Long-term Lightcurve and Volatile-driven Activity}
\label{s.lightcurve}
\label{s.activity}
Due to the density and slow crossing time of dust within 2I's coma at the scale of our ground-based observations as discussed in \citet[][]{Jewitt2019}, measuring any short term lightcurve variations on the order of hours to 10's of hours caused by the rotation of the comet's nucleus is difficult. However, other effects on the comet's brightness can happen on longer time spans of weeks to months such as outbursts, seasonal effects or changes in its activity due to the sublimation of different volatile species that become active at different heliocentric distances 
along the comet's orbit \citep[][]{Hughes1990, Li2016, Keller2017,Womack2017}. Because these effects can take weeks to months, a comet needs to be monitored over a long-time period requiring the dedication of observers to make regular observations of the comet.  A detailed discussion of the long-term lightcurve's implication for the activity of 2I follows.

\begin{figure*}
\centering
\includegraphics[width=1.0\textwidth]{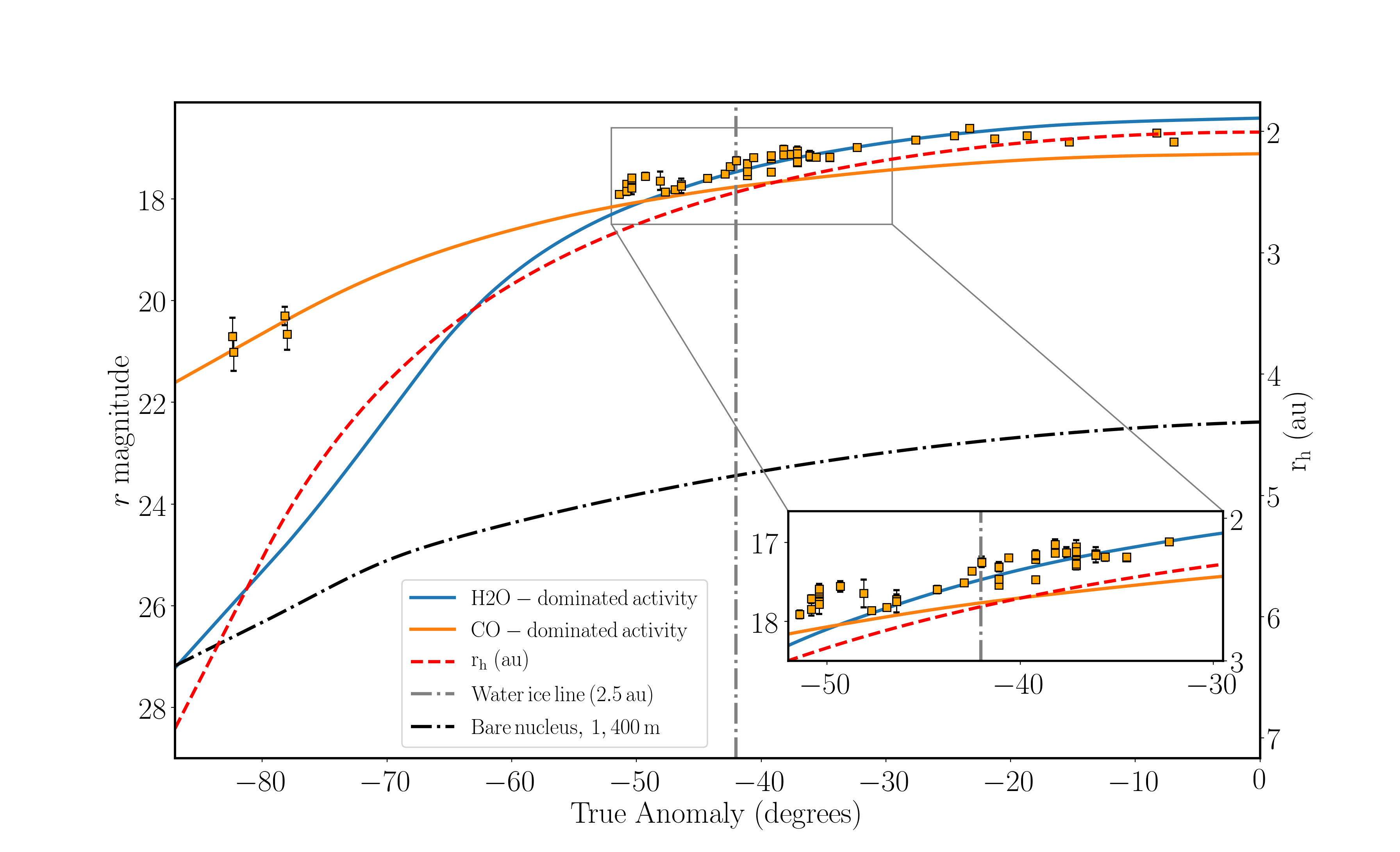}
\caption{$r$ magnitude of 2I as a function of the true anomaly using photometry translated to $r$ magnitudes for data taken between 2019 March 17 \citep{Ye2019b} and 2019 November 29 (this campaign), and tabulated in Table~\ref{tab:phot}. The blue and orange lines are the predicted brightness as a function of true anomaly angle for H$_2$O and CO-dominated activity for a comet with a diameter of 1.4 km and 100$\%$ active surface area from the out-gassing model of 2I from \citep[][]{Fitzsimmons2019}. The brightness prediction assumes a 5\arcsec~aperture, comparable to the aperture size used to measure the brightness of 2I in this study. The dash-dot black line is the predicted brightness as a function of true anomaly angle assuming an inactive, bare nucleus, 1.4 km diameter and 0.04 albedo, the lower limit on the estimate of 2I's nucleus size from the detection of CN gas \citep[][]{Fitzsimmons2019}. The red-dashed line shows the heliocentric distance, $r_h$ as a function of true-anomaly for 2I. The vertical, grey dash-dot line is positioned on the true anomaly where 2I crosses the water ice line at 2.5 au. True anomaly = 0$^{\circ}$ corresponds to 2I's perihelion passage on 2019 December 08 UTC.}
\label{fFiig:brightnessvtime}
\end{figure*}

The lightcurve of equivalent $r$ magnitudes is plotted in Fig.~\ref{fFiig:brightnessvtime}. Currently as of writing, brightness of 2I, plotted as orange squares, appears to follow the trend in predicted by \citep[][]{Fitzsimmons2019} for a H$_2$O-dominated comet plotted as a blue line, best seen in the inset plot zoomed in on $-$52$^{\circ}$ to $-$32$^{\circ}$ in Fig.~\ref{fFiig:brightnessvtime}. The activity for a CO$_2$ dominated comet is plotted as a solid orange line. Both of these activity models from \citet[][]{Fitzsimmons2019} are based on assuming a nucleus diameter of $\sim$1 km and activity consistent with Solar System comets using the measured CN activity to estimate the production rate of other volatile species \citep[][]{AHearn1995}. In addition, the activity model assumes 100$\%$ of the comet's surface is active and and dust grain properties similar to Solar System comets. There is a recent rise in brightness as 2I approached the water ice line at heliocentric distance, $r_h$ = 2.5 au on 2019 October 02 UTC that may correspond to the increase in the sublimation rate of H$_2$O as the comet approaches the Sun \citep[][]{Meech2004,Jewitt2015a}.

Extrapolating the H$_2$O brightness model backwards to the pre-discovery data taken by ZTF in 2019 March and 2019 May when the comet was at a heliocentric distance of 6.03 au and 5.09 au respectively, predicts a much fainter magnitude of $r\sim$26 than the observed magnitude of $r$ = 20.5 to 21.0. As shown by \citep[][]{Ye2019b} and confirmed by our work, the actual observed pre-discovery $r$ magnitudes are much closer to the brightness model predicted for a comet that has its activity dominated by CO than H$_2$O \citep[][]{Fitzsimmons2019}. This is supported by the fact that H$_2$O is very weakly sublimating at temperatures $\lesssim$150 K at a heliocentric distance $>$3.5 au while CO can become volatile much further from the Sun at heliocentric distances exceding 10 to 100 au \citep[][]{Meech2004}. 

However, the pre-discovery photometry may also be compatible with CO$_2$-driven activity where CO$_2$ can become active at $>13$ au \citep[][]{Womack2017, Ye2019b}. As discussed in Section~\ref{sec:massloss}, a significant production rate of H$_2$O is inferred from the observed production of CN and C$_2$ gas \citep[][]{Fitzsimmons2019,Kareta2019} and is $\sim$100 kg/s comparable, though larger than the H$_2$O $\sim$20 kg/s production inferred from the detection of the [O I] 6300 $\mathrm{\AA}$ line taken at further heliocentric distances \citep[][]{McKay2019}. Since our photometric lightcurve suggests that the activity of 2I is partially driven by CO, we expect the mass loss of CO to also be much higher than the mass loss from dust in the $\sim$10-100 kg/s range as it approaches perihelion as the ratio of CO to H$_2$O is has been shown to be $>$130$\%$ as revealed by recent \textit{Hubble Space Telescope} (\textit{HST}) and Alma observations \citep[][]{Cordiner2020, Bodewits2020}, much higher than the typical $<$30$\%$ of Solar System comets \citep[][]{Paganini2014,Meech2017b}.

The difference between the observed brightness of 2I in the pre-discovery data is even larger for a bare, inactive $\sim$1.4 km diameter nucleus as seen for the black dash-dot line in Fig.~\ref{fFiig:brightnessvtime}. In addition, there appears to be a $\sim$0.2 mag change in brightness in the lightcurve between 2019 September 20 and 2019 October 03 UTC corresponding to true anomaly angles $-$47$^{\circ}$ and $-$42$^{\circ}$ deviating from the trend predicting the brightness for a H$_2$O dominated comet as seen in Fig.~\ref{fFiig:brightnessvtime} possibly indicating a change in the activity of the comet. We must caution that the height of the curves is also dependent on the size of the nucleus and the activity could be compatible with a slight increase in nucleus size and a corresponding decrease in the water production rate.

Concerning the out-gassing models used to constrain the activity, it is important to note that our suggestion of initial CO driven outgassing activity transitioning to H$_2$O driven activity is not dependent on 2I's nucleus size or fractional active out-gassing area. The fractional active outgassing scales the CO + H$_2$O model; once set, this scale is fixed. It is the relative shape of our measured 2I long term lightcurve, and the upward inflection point in the lightcurve seen at distances r$_h$ $< $3 au, that tell us that additional water outgassing has turned on and thence started to dominate the activity of the object. This latter finding, of water outgassing dominance, again tells us that 2I appears to be acting like a normal solar system comet, as water is by far the most abundant ice found in solar system comets.

\subsection{Mass Loss}
\label{sec:massloss}
Using the $g$, $r$, $i$ and $z$ photometry obtained by the ARC 3.5 m on 2019 September 12 UTC, we place estimates on 2I's $Af\rho$ parameter, a proxy for dust production rate \citep[][]{AHearn1984}. We find $(Af\rho)_g$ = 113 $\pm$ 5 cm, $(Af\rho)_r$ = 185 $\pm$ 7 cm, $(Af\rho)_i$ = 223 $\pm$ 8 cm and $(Af\rho)_z$ = 180 $\pm$ 8 cm, typical values for Solar System comets \citep[][]{AHearn1995,Kelley2013}, implying an out-gassing rate of $\sim$10$^{27}$ mol/s \citep[][]{Fink2012}. The recently taken data from between 2019 September 11 UTC and 2019 December 20 UTC seen in Table~\ref{tab:phot} and Fig.~\ref{fFiig:brightnessvtime} shows a brightening trend of $\sim$0.03 mag/day consistent with the enhancement in brightness expected for the evolving viewing geometry of the comet according to the following equation
\begin{equation}
\label{eqn.brightness}
m_V = H_{abs} + 2.5 \mathrm{log_{10}}(r_h \Delta) + \Phi(\alpha)
\end{equation}
where $m_V$ is the apparent magnitude, $H_{abs}$ is the absolute magnitude, $r_h$ is the heliocentric distance in au, $\Delta$ is the observer-centric distance in au, $\Phi(\alpha)$ is a function describing the brightening of the comet which we approximate with $\Phi(\alpha)$ = $-$0.04$\alpha$ \citep[][]{Jewitt1991} and $\alpha$ is the phase angle of the comet measured in degrees, appropriate for comets at smaller phase angles than $\sim$20$^{\circ}$ \citep[][]{Bertini2017}. 
We translate the $H_{abs}$ magnitude computed from Eq.~\ref{eqn.brightness} into an effective cross-section, $C$, in units of km$^2$ within a 10,000 km aperture using the following formula
\begin{equation}
\label{eqn.crosssetion}
C = 1.5 \times 10^{6}\, p_v^{-1} \, 10^{-0.4H}
\end{equation}
from \citet[][]{Jewitt2016}, where $p_v$ is the albedo of the comet, assumed to be 0.10, typical for comet dust \citep[][]{Jewitt1986, Kolokolova2004}. We caution that uncertainties of $H_{abs}$ inferred from Eq.~\ref{eqn.brightness} are lower limits on the overall photometric uncertainty because they should also include a component from the phase function which is unknown at the present time for 2I. 

\begin{figure*}
\centering
\includegraphics[width=0.95\textwidth]{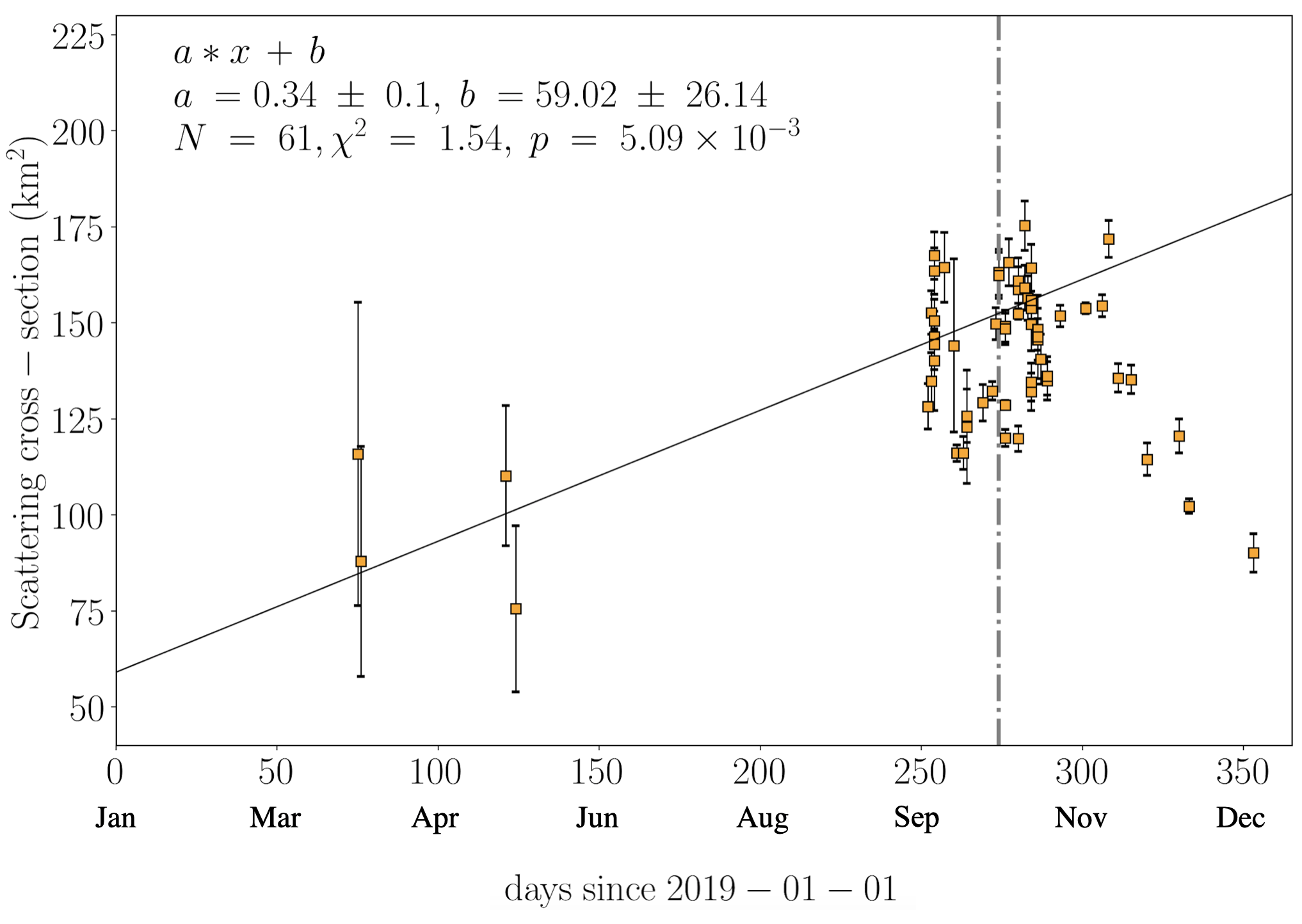}
\caption{The effective cross-section of 2I calculated from Eq.~\ref{eqn.crosssetion} as a function of days since 2019-01-01 UTC. The black line shows the minimized $\chi^2$ fit to the cross-section measurements and the vertical dash-dot line corresponds to the date when 2I crossed the water-ice line at 2.5 au.}
\label{fFiig:crosssectionvstime}
\end{figure*}

We plot the effective cross-section over the baseline of available 2I photometry including the ZTF pre-discovery data taken in March and May 2019 as seen in Fig.~\ref{fFiig:crosssectionvstime}. The median cross-section from these data is $\sim$145 km$^2$. A linear fit is applied to the data with the minimized $\chi^2$ fit corresponding to slope of 0.34 $\pm$ 0.10 km$^{-2}$/day suggesting that the cross-section doubled since the earliest observations from the ZTF pre-discovery images in 2019 March 17 and will exceed $\sim$200 km$^2$ by the time 2I reaches perihelion on 2019 December 08 UTC assuming the slope is constant. We note that the data point corresponding to the 2019 November 29 UTC and 2019 December 20 UTC data may be due to 2I increasing in brightness at a slower rate than expected as the comet reaches perihelion, so we do not include it with our linear fit.

There appears to be a sudden, $\sim$50 km$^2$ jump in the effective cross-section between 2019 September 20 and 2019 October 03 UTC as seen in Fig.~\ref{fFiig:crosssectionvstime} corresponding to the drop in the overall trend for brightness seen in the light-curve plotted in Fig.~\ref{fFiig:brightnessvtime} for between true anomaly angles $-$47$^{\circ}$ and $-$42$^{\circ}$. As discussed above in Section~\ref{s.lightcurve}, the deviation in brightness may indicate a change in the comet's activity. The location of the vertical dashed grey line in Fig.~\ref{fFiig:crosssectionvstime} indicates when 2I crossed the water ice line which is nearby to an observed steep increase in the cross-section, possibly connected to the sublimation of H$_2$O discussed further in Section~\ref{s.activity}. There is also another, earlier $\sim$50 km$^2$ jump in the cross-section starting around the onset of our observations on 2019 September 10 UTC, though we caution that the variability can also be due to the large errors of the individual data points.

\subsection{Diameter Estimate}
\label{s.diameter}
A rough upper-limit to the diameter of 2I of $\sim$5-10 km was found using our conventional ground-based observations, typically on the order of $\sim$1\arcsec~resolution similar to the size upper limit estimate of $\sim$8 km from \citet[][]{Jewitt2019}. Coma-subtraction techniques that remove the dust component from the total effective cross-section of the comet \citep[ie.,][]{Fernandez2013,Bauer2017}, proved to be only partially effective due to the density of the coma at the resolution afforded by ground-based observations.

A more accurate upper limit can be inferred by measuring the effective cross-section using high resolution data from high resolution ground-based AO and space-based observations from Keck \citep[e.g.,][]{Marchis2006}. Using a 0.48\arcsec~aperture with a contiguous median sky-subtraction annuli from 0.48\arcsec~to 0.96\arcsec, we obtain $Kp$ = 19.63 $\pm$ 0.09. We use our visible and NIR colors determined for 2I to transform the $K_p$ magnitude measured in the OSIRIS images taken on 2019 October 04 UTC to $V$ = 21.95 $\pm$ 0.16 from our combined VisNIR photometry and IRTF spectrum presented in  S~\ref{sec:photometry} and \ref{s.nir}. We use the $V$ magnitude to calculate an $H_{abs}$ = 16.88 $\pm$ 0.16 using Eq.~\ref{eqn.brightness} and with a $r_h$ = 2.48 au, $\Delta$ = 2.96 au and $\alpha$ = 18.65$^{\circ}$ that the comet had on 2019 October 04 UTC. As mentioned above in Section~\ref{sec:massloss}, the uncertainty on the $H_{abs}$ calculation is a lower limit due to the unknown phase function of the comet. 

We converted the $H_{abs}$ magnitude determined with the 0.48\arcsec~aperture into an effective cross-section using Eq.~\ref{eqn.crosssetion} resulting in an effective cross-section of 2.65 $\pm$ 0.39 km$^2$ assuming an albedo equal to 0.1, typical for comet dust and resulting in a value of 6.63 $\pm$ 0.97 km$^2$, assuming an albedo equal to 0.04, typical for comet nuclei \citep[][]{Fernandez2001,Bauer2017}. A higher albedo could also be used to calculate the cross-section corresponding to an icy, more reflective composition \citep[][]{Yang2009}, but the NIR spectra presented here as well as additional NIR spectra \citep[][]{Yang2019} do not show strong evidence for presence of ice in the coma of 2I.

Using the following equation to calculate the diameter from $C$, $D = 2 \sqrt{C/\pi}$, we obtain the values 1.84 $\pm$ 0.13 km and 2.90 $\pm$ 0.21 km for $pv$ = 0.1 and 0.04 respectively, implying a mass of $\lesssim$10$^{12}$ kg assuming a comet nucleus density of 400 kg/m$^3$ \citep[e.g.,][]{Patzold2016}. In addition to the advantages of using higher resolution AO imaging compared to conventional ground-based observations, observing comets in longer wavelengths such as $K_p$ band has the advantage of avoiding much of the scattered light from micron-sized dust that is more prevalent in visible wavelengths. This effect of using less dust-contaminated wavelengths in photometry of comets has already been demonstrated to produce robust diameter estimates of comets even at spatial resolutions approaching or worse than in the $K_p$ AO images presented here \citep[][]{Fernandez2013,Bauer2017}.

We caution that the estimates of the nucleus size are strictly rough upper limits. Profiles through the imagery, especially the high spatial resolution Keck images, do not show a discernible signal due to a point source nucleus arguing for an object dominated in brightness by scattered light from its surrounding coma \citep[e.g.,][]{Jewitt2019,Kim2020,Bolin2020hst}, and suggesting a small (less than a few km diameter nucleus) at the ~2.9 au distance 2I was observed at by Keck.

We thus resort to estimating its nucleus size in 2 different ways: (1) a very optimistic method that includes all the flux detected in the central point spread function (PSF), in order to determine a hard upper limit for the nucleus' size; (2) a more realistic method that involves extrapolating the run of coma brightness versus distance from the nucleus into the central PSF, allowing us to model the coma in the entire image and thence remove it, and (3) a hybrid approach whereby we take the flux from method (1) and modify it for known observations of hyperactive solar system comets. 

The first method yields an object with diameter of  $\sim$3 km, giving us a hard upper limit to 2I's size - it can't be on the order of 20 km diameter or greater, as some initial estimates have stated. The second method is much more constraining, as we do not detect a nucleus residual after modeling and removing the coma (assuming a stellar PSF derived from cuts though highly trailed stars perpendicular to the trailing direction). Adopting a 2 $\sigma$ upper limit from the noise level of the coma removal ($\sim$10$\%$ of the central PSF flux), we find an upper limit to Borisov diameter of $\sim$1.4 km similar to the prediction of 2I's size by the thermal model presented in \citet[][]{Fitzsimmons2019}. The third method takes note of the fact that a small Borisov nucleus size implies a very high outgassing rate per unit km$^2$ of nucleus surface area, a phenomenon seen for ``hyperactive'' solar system comets like 103P Hartley 2 \citep[][]{Lisse2009, AHearn2011, Harker2018} and 46P/Wirtannen \citep[][]{Lis2019} to be due to large amounts of ice-rich dust expulsion into the surrounding coma, greatly increasing the active surface area receiving solar insolation. Using the ratio of $\sim$4:1 coma: nucleus surface brightness seen for comet 103P during the deep impact mission in situ flyby, we can scale the total flux in the central PSF by a factor of 1/(1+4) = 0.2, and then proceed as if we have measured the nucleus' flux. Doing so we arrive again at an estimated nucleus diameter upper limit of $\sim$1.4 km similar to nucleus measurements from high-resolution, space-based observations \citep[][]{Jewitt2019hst}.

\section{Discussion and Conclusions}
The second interstellar object, 2I, seems on all accounts like an ordinary comet compared to the comets of the Solar System, though it is depleted in some chemical species relative to Solar System comets \citep[][]{Opitom2019,Kareta2019,Bannister2020} and has an excess of CO \citep[][]{Bodewits2020,Cordiner2020}. If it were not for its significantly hyperbolic orbit, 2I probably would not have warranted an in-depth scientific investigation. However, given its special status as a comet of extra-solar origin, it presents a unique opportunity to study the cometary components of other star systems since a likely outcome of the evolution of planetary systems is the ejection of many cometary bodies \citep[][]{Raymond2018a,Raymond2018b}. In our own Solar System, the comet population is a record of its formation properties and evolution \citep{Morbidelli2019}, so by studying objects that were ejected from their home systems like 1I and 2I, we can directly observe the consequences of planetary system evolution.

One of the salient properties of 2I is that it contains significant amounts of volatiles such as CN and C$_2$ gas \citep[][]{Fitzsimmons2019,Kareta2019, Opitom2019} and there is evidence in this work from the photometry presented in Section~\ref{s.lightcurve} that the comet also contains H$_2$O unlike the super rich CO/N2/CH4, H$_2$O depleted comet C/2016 R2 \citep[][]{Cochran2018,McKay2019}. Instead, it is acting like an Oort Cloud comet on a Myr-period orbit like C/1995 O1 (Hale-Bopp) or C/2013 S1 (ISON) or C/2017 K2 \citep[][]{Jewitt2017cc, Meech2017}, which commonly demonstrate out-gassing abundances of CO with respect to water in the 0.2 $-$ 20$\%$ range \citep[][]{BockeleeMorvan2004}. The presence of moderately abundant CO and H$_2$O on 2I \citep[][and this work]{Ye2019b} suggests that while 2I hasn't been heated so thoroughly by its home Sun (as Solar System Jupiter Family comets and likely 1I have), it could have been ejected from its home system or was placed into its star's equivalent of the Solar System's Oort Cloud more than a few Myrs of its formation after its home system's protoplanetary disk midplane had cleared enough to heat its surface above 30 K \citep[][]{Lisse2019}. This assumes that in comparison with the Solar System comet C/2016 R2 has never been heated above 20K before encountering the Sun, where it is in the process of losing its hypervolatiles, but not its H$_2$O ice due to hypervolatile supercooling \citep[][]{Biver2018, Lisse2019}.  Additionally, the host star of 2I may have a higher stellar iron abundance that has been shown to have an effect on the water ice fraction solid building blocks in the protoplanetary disk favoring a higher concentration of CO/CO$_2$ relative to water ice \citep[][]{Bitsch2020}.

Compared to the 2I, 1I had only marginal levels of activity. The activity of 1I was not seen in direct imaging of the comet or in its spectra \citep[][]{Meech2017,Fitzsimmons2018}, only being evident via detailed astrometry of the small trajectory deviations from inertial-solar gravitational caused by low levels of out-gassing \citep[][]{Micheli2018}. So if it was actively out-gassing, its coma was very faint and below the noise level in any of the detection images, including imaging from \textit{HST}. One explanation for the lack of activity of 1I is that it had a mantle built up by cosmic ray bombardment during its interstellar travel, trapping its volatiles inside its structure \citep[][]{Fitzsimmons2017}. On the other hand, the specific out-gassing rate per unit body surface area implied by the non-graviational force model of \citet[][]{Micheli2018} is on the upper-bound of Jupiter family comet activity \citep[][]{Fernandez2013}. 1I was small, on $\sim$250 m diameter \citep[][]{Meech2017, Trilling2018} compared to the typical km scale for a JFC comet, that it took very little force from out-gassing to significantly accelerate it. 

The activity of 2I can possibly be used to distinguish between the ``large'' or ``small'' size estimates for 2I discussed in Section~\ref{s.diameter}, especially in comparison versus 1I, by constraining the effect of non-gravitational forces due to outgassing on its orbital trajectory. Moderate non-graviational force parameters have been measured for the orbit of 2I in pre-discovery data when the comet's activity was weaker \citep[][]{Ye2019b} as has been done for Solar System comet \citep[e.g.,][]{Moreno2017}. If 2I has a similar size as 1I, then its small total volume and mass means that it could also be substantially accelerated much more by non-gravitationally outgassing jet forces compared to 1I given the apparent much larger outgassing rate for 2I than 1I. However if 2I is much larger than 1I where the mass ratio between 2I and 1I scales as  (3 km /0.25 km)$^3$, $\sim$ 1,000 times more massive than 1I, then 2I can be outgassing $\sim$ 1000 times more than 1I and still suffer the same amount of jet acceleration. Thus monitoring the astrometric position of 2I throughout the next few months will be critical for understanding the size regime of 2I's nucleus as its activity grows and its orbit can be more potentially affected by non-gravitational forces.

Other estimates of size distribution for the ISO population have included both upper limits on the non-detection of ISOs \citep[][]{Engelhardt2017} and on the sole detection of 1I \citep[][]{Trilling2018a, Raymond2018a}. We estimate the size distribution of the ISO population updated with the detection of 2I and the upper limit on its diameter from high-resolution images. We calculate the number of 250 m ISOs to be $\sim$ 13 objects within 3 au of the Sun by scaling the density of 250 m ISOs, $\sim$1 ISO within 1 au of the Sun at any given time, \citep[][]{Meech2017} to a sphere of radius 3 au accounting for the gravitational focusing of the Sun assuming a velocity at infinity of 32km/s for ISOs as for comet 2I. Assuming a slightly lower velocity at infinity of 26 m/s as for 1I does not significantly change the results. 

\begin{figure*}
\centering
\includegraphics[width=0.75\textwidth]{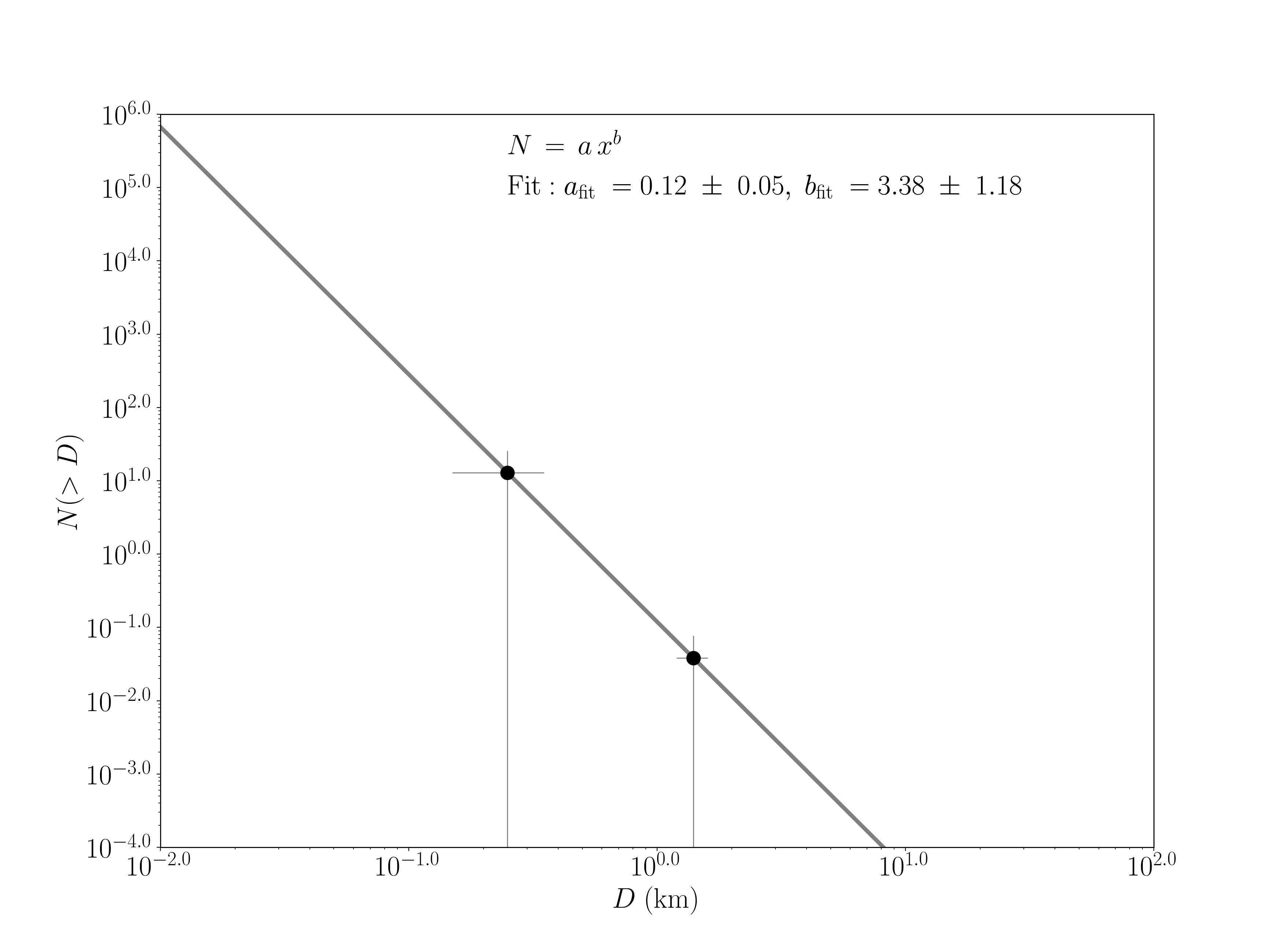}
\caption{The size distribution of ISOs within 3 au of the Sun estimated from the detection of 1I with D$\simeq$250 m and 2I with D $\sim$1.4 km. The number of ISOs in the size range of 1I is estimated to be $\sim$340 from the rate of occurrence of 1I-sized objects. The solid grey line is fit to data with the function $y \, = \, ax^b$ and is based on the estimated size of 1I from the literature and the average of the upper limits on the diameter of 2I assuming 0.04 and 0.1 albedo. The error bars on the number of 1I and 2I objects are estimated to be $\sim$10$^{-2}$ and $\sqrt{13}$ respectively. The errors on the parameters are determined with the allowable range by the errors the diameter of 1I and 2I \citep[][,this work]{Trilling2018} and number of 1I-like and 2I-like objects assuming Poissonian statistics}.
\label{fFiig:ISOSFD}
\end{figure*}

We calculate the relative number of 250 m diameter ISOs like 1I to 1.4 km ISOs like 2I by the fraction of time 2I was observable within 3 au over the total survey lifetime of the past 15 y. We consider 15 y the amount of time that the search for objects like 2I by amateur astronomers to be active due to the difficulty in obtaining sensitive CCD cameras at the consumer level before this time \citep[][]{Copandean2019}. This translates into 13$\pm$13 ISOs with diameter $\sim$250 m to $\sim$4$\times$10$^{-2}$$\pm$4$\times$10$^{-2}$ ISO in with diameter $\sim$2-3 km within 3 au of the Sun where the uncertainties are estimated from the allowable range in number of 1I-like and 2I-like objects assuming Poissonian statistics. The resulting cumulative size distribution inferred from the ratio of the number of 250 m objects to 1.5-3 km objects is shown in Fig.~\ref{fFiig:ISOSFD}. The slope of the cumulative size distribution is $\sim$-3.38 $\pm$ 1.18, which is comparable to the cumulative size distribution slope of collisionally evolved Solar System bodies \citep[][]{Dohnanyi1969} and of comets measured in the km diameter range \citep[][]{Meech2004,Fernandez2013,Boe2019}.

The size estimates we have derived for 2I above bear on the question of why ``asteroidal'' object 1I/Oumuamua was detected before an active, bright object like cometary 2I. Cognizant of the dangers of extrapolating size distributions and population statistics from a sample of N=2 purportedly related objects, we do so here because these arguments will likely be valuable in the fullness of time as we collect more and more detections of interstellar objects over the next decades. Naively one would have expected active, $\sim$3 magnitudes brighter 2I-like objects to have been detected first modulo selection effects \citep[e.g.,][]{Jedicke2016,Vokrouhlicky2017a} because they can be seen out to much farther distances (the detectability distance scales as the object's $D$, so the volume of space it can be detected in goes as the objects $D^2$). If 2I is substantially bigger than 1I, then for a steep enough interstellar object size distribution, (slope steeper than -3), there can be many more 1Is in the volume of space than 2I-like objects, enough so that 1I-like objects will be seen more frequently. For a size distribution scaling with $\sim D^{-3}$ and $D_{1I} \simeq$250 m, $D_{2I}\simeq$ 2.0 km, there would be several 100s of 1I-like objects for every 1 Borisov in a given volume of space, overwhelming the 100 times larger volume that a 2 km diameter 2I-like object could be detected in.

Recent evidence suggests that the slope of active comets goes from steeper to shallower at a transition boundary of $D\sim$3 km \citep[see Fig.~5 of][]{Boe2019}. In the $D\lesssim$3 range, the cumulative size distribution slope is significantly shallower than objects $\gtrsim$3 km in size which seems to contradict the slope of the size distribution that we measure for the ISOs. However, work in preparation on the size distribution of inactive comets in the sub-km range shows a steeper size distribution than compared to sub-km active comets, and more closely resembles the slope of the ISO size distribution from this work \citep[][]{Boe2019b}.  Assuming the properties of the size distributions of active and inactive comets in the Solar System are shared with those in extra-solar systems, 1I may be a representative of the inactive comet population given its lack of activity and may come from a population with a steep size distribution explaining its small size. The observed activity of 2I suggests that it comes from an active comet population which has a shallower size distribution that might explain the case if 2I has a larger compared to 1I. Alternatively, if 2I actually has a size comparable to 1I, its apparent activity might indicated that it comes from a shallower size distribution compared to the size distribution of inactive objects. Thus, the relative numbers of active and inactive objects from correspondingly shallower and steeper size distributions may explain the relative frequency of inactive objects like 1I compared to inactive objects.

The ISO size distribution that we are observing may be a hybrid of both active and inactive comet populations from the ensemble of comet-ejecting extra-solar systems producing a mixed active and inactive observed ISO population resulting in a size distribution slope steeper than $\sim$-3. Although the error bars from our measurement of the ISO size distribution slope from the occurrence of 1I and 2I are large, future observations of ISOs could refine the measurement of the slope. A shallower slope would be more consistent with production from a population in collisional equilibrium \citep[e.g.,][]{Dohnanyi1969} while a steeper slope may indicate that the ISO population is fed partially by additional fragmentation events such as from tidal disruption \citep[][]{Bolin2018,Raymond2018bc,Zhang2020}. In any case, the existence of sub-km interstellar comets like 1I suggests that the size distribution of objects in extra-solar Kuiper Belts, the progenitors of extra-Solar comets, is not truncated at 1-2 km challenging the claim that the size distribution of objects in the Solar System's Kuiper Belt is effectively truncated at 1-2km in diameter \citep[][]{Singer2019}. The arrival of additional ISOs will provide further constraints on their physical properties and size distribution enhancing our understanding of comets in extra-solar systems. 

\acknowledgments

We would like to thank the anonymous reviewers for their helpful comments in substantially improving the manuscript.

This work was supported by the GROWTH project funded by the National Science Foundation under PIRE Grant No 1545949.

Our work includes observations obtained with the Apache Point Observatory 3.5-meter telescope, which is owned and operated by the Astrophysical Research Consortium. We thank the Director (Nancy Chanover) and Deputy Director (Ben Williams) of the Astrophysical Research Consortium (ARC) 3.5m telescope at Apache Point Observatory for their enthusiastic and timely support of our Director's Discretionary Time (DDT) proposals. We also thank Russet McMillan, Ted Rudyk, Candace Gray, Jack Dembicky and the rest of the APO technical staff for their assistance in performing the observations on the same day our DDT proposals were submitted.

Based on observations obtained with the Samuel Oschin Telescope 48-inch and the 60-inch Telescope at the Palomar Observatory as part of the Zwicky Transient Facility project. ZTF is supported by the National Science Foundation under Grant No. AST-1440341 and a collaboration including Caltech, IPAC, the Weizmann Institute for Science, the Oskar Klein Center at Stockholm University, the University of Maryland, the University of Washington, Deutsches Elektronen-Synchrotron and Humboldt University, Los Alamos National Laboratories, the TANGO Consortium of Taiwan, the University of Wisconsin at Milwaukee, and Lawrence Berkeley National Laboratories. Operations are conducted by COO, IPAC, and UW.

Some of the data presented herein were obtained at the W. M. Keck Observatory, which is operated as a scientific partnership among the California Institute of Technology, the University of California and the National Aeronautics and Space Administration. The Observatory was made possible by the generous financial support of the W. M. Keck Foundation.

We thank Jim Lyke and Carlos Alvarez for guiding the planning and executing the adaptive optics observation with the OSIRIS instrument on Keck I.

C-C Ngeow thanks the funding from MoST grant 104-2923-M-008-004-MY5.

The work of DS was carried out at the Jet Propulsion Laboratory,
California Institute of Technology, under a contract with NASA.

SED Machine is based upon work supported by the National Science Foundation under Grant No. 1106171

This publication has made use of data collected at Lulin Observatory, partly supported by MoST grant 108-2112-M-008-001.

The results presented in this paper are based in part on observations collected with the Liverpool Telescope, which is operated on the island of La Palma by Liverpool John Moores University in the Spanish Observatorio del Roque de los Muchachos of the Instituto de Astrofisica de Canarias with financial support from the UK Science and Technology Facilities Council.

Visiting Astronomer at the Infrared Telescope Facility, which is operated by the University of Hawaii under contract NNH14CK55B with the National Aeronautics and Space Administration.

The authors wish to recognize and acknowledge the very significant cultural role and reverence that the summit of Maunakea has always had within the indigenous Hawaiian community. We are most fortunate to have the opportunity to conduct observations from this mountain.

FM is supported by an appointment to the NASA Postdoctoral Program at the Jet Propulsion Laboratory, administered by Universities Space Research Association under contract with NASA.

The authors would like to thank E. Turner, B Draine, S. Tremaine, and M Mac-Low for many useful discussions concerning the nature and provenance of 1I and 2I. In addition, the authors would like to thank R. Jedicke and G. Helou for helpful discussion on the size distribution of comets. We would also like to thank J. Bauer and Y. Fernandez for fruitful discussion on comet nuclei. 

The authors would like to thank the astrophysics masters students of the Universit\'{e} C\^{o}te d'Azur who recorded the data taken by the C2PU telescope at Observatoire de la C\^{o}te d'Azur, Calern observing site.

\facility{Astrophysical Research Consortium 3.5 m, C2PU Omicron telescope, Bisei Observatory 101 cm, Lulin Optical Telescope, Keck:I (LRIS, OSIRIS), Liverpool Telescope, Mount Laguna Observatory 40-inch, NASA/Infrared Telescope Facility, Zwicky Transient Facility, SED Machine}

\bibliographystyle{aasjournal}
\bibliography{ms}

\begin{thebibliography}{}
\expandafter\ifx\csname natexlab\endcsname\relax\def\natexlab#1{#1}\fi
\providecommand{\url}[1]{\href{#1}{#1}}
\providecommand{\dodoi}[1]{doi:~\href{http://doi.org/#1}{\nolinkurl{#1}}}
\providecommand{\doeprint}[1]{\href{http://ascl.net/#1}{\nolinkurl{http://ascl.net/#1}}}
\providecommand{\doarXiv}[1]{\href{https://arxiv.org/abs/#1}{\nolinkurl{https://arxiv.org/abs/#1}}}

\bibitem[{{A'Hearn} {et~al.}(1995){A'Hearn}, {Millis}, {Schleicher}, {Osip}, \&
  {Birch}}]{AHearn1995}
{A'Hearn}, M.~F., {Millis}, R.~C., {Schleicher}, D.~O., {Osip}, D.~J., \&
  {Birch}, P.~V. 1995, \icarus, 118, 223, \dodoi{10.1006/icar.1995.1190}

\bibitem[{{A'Hearn} {et~al.}(1984){A'Hearn}, {Schleicher}, {Millis}, {Feldman},
  \& {Thompson}}]{AHearn1984}
{A'Hearn}, M.~F., {Schleicher}, D.~G., {Millis}, R.~L., {Feldman}, P.~D., \&
  {Thompson}, D.~T. 1984, \aj, 89, 579, \dodoi{10.1086/113552}

\bibitem[{{A'Hearn} {et~al.}(2011){A'Hearn}, {Belton}, {Delamere}, {Feaga},
  {Hampton}, {Kissel}, {Klaasen}, {McFadden}, {Meech}, {Melosh}, {Schultz},
  {Sunshine}, {Thomas}, {Veverka}, {Wellnitz}, {Yeomans}, {Besse}, {Bodewits},
  {Bowling}, {Carcich}, {Collins}, {Farnham}, {Groussin}, {Hermalyn}, {Kelley},
  {Kelley}, {Li}, {Lindler}, {Lisse}, {McLaughlin}, {Merlin}, {Protopapa},
  {Richardson}, \& {Williams}}]{AHearn2011}
{A'Hearn}, M.~F., {Belton}, M. J.~S., {Delamere}, W.~A., {et~al.} 2011,
  Science, 332, 1396, \dodoi{10.1126/science.1204054}

\bibitem[{{Bannister} {et~al.}(2017){Bannister}, {Schwamb}, {Fraser},
  {Marsset}, {Fitzsimmons}, {Benecchi}, {Lacerda}, {Pike}, {Kavelaars},
  {Smith}, {Stewart}, {Wang}, \& {Lehner}}]{Bannister2017}
{Bannister}, M.~T., {Schwamb}, M.~E., {Fraser}, W.~C., {et~al.} 2017, \apjl,
  851, L38, \dodoi{10.3847/2041-8213/aaa07c}

\bibitem[{{Bannister} {et~al.}(2020){Bannister}, {Opitom}, {Fitzsimmons},
  {Moulane}, {Jehin}, {Seligman}, {Rousselot}, {Knight}, {Marsset}, {Schwamb},
  {Guilbert-Lepoutre}, {Jorda}, {Vernazza}, \& {Benkhaldoun}}]{Bannister2020}
{Bannister}, M.~T., {Opitom}, C., {Fitzsimmons}, A., {et~al.} 2020, arXiv
  e-prints, arXiv:2001.11605.
\newblock \doarXiv{2001.11605}

\bibitem[{{Bauer} {et~al.}(2017){Bauer}, {Grav}, {Fern{\'a}ndez}, {Mainzer},
  {Kramer}, {Masiero}, {Spahr}, {Nugent}, {Stevenson}, {Meech}, {Cutri},
  {Lisse}, {Walker}, {Dailey}, {Rosser}, {Krings}, {Ruecker}, {Wright}, \& {the
  NEOWISE Team}}]{Bauer2017}
{Bauer}, J.~M., {Grav}, T., {Fern{\'a}ndez}, Y.~R., {et~al.} 2017, \aj, 154,
  53, \dodoi{10.3847/1538-3881/aa72df}

\bibitem[{{Bellm} {et~al.}(2019{\natexlab{a}}){Bellm}, {Kulkarni}, {Graham},
  {Dekany}, {Smith}, {Riddle}, {Masci}, {Helou}, {Prince}, {Adams},
  {Barbarino}, {Barlow}, {Bauer}, {Beck}, {Belicki}, {Biswas}, {Blagorodnova},
  {Bodewits}, {Bolin}, {Brinnel}, {Brooke}, {Bue}, {Bulla}, {Burruss}, {Cenko},
  {Chang}, {Connolly}, {Coughlin}, {Cromer}, {Cunningham}, {De}, {Delacroix},
  {Desai}, {Duev}, {Eadie}, {Farnham}, {Feeney}, {Feindt}, {Flynn},
  {Franckowiak}, {Frederick}, {Fremling}, {Gal-Yam}, {Gezari}, {Giomi},
  {Goldstein}, {Golkhou}, {Goobar}, {Groom}, {Hacopians}, {Hale}, {Henning},
  {Ho}, {Hover}, {Howell}, {Hung}, {Huppenkothen}, {Imel}, {Ip}, {Ivezi{\'c}},
  {Jackson}, {Jones}, {Juric}, {Kasliwal}, {Kaspi}, {Kaye}, {Kelley},
  {Kowalski}, {Kramer}, {Kupfer}, {Landry}, {Laher}, {Lee}, {Lin}, {Lin},
  {Lunnan}, {Giomi}, {Mahabal}, {Mao}, {Miller}, {Monkewitz}, {Murphy},
  {Ngeow}, {Nordin}, {Nugent}, {Ofek}, {Patterson}, {Penprase}, {Porter},
  {Rauch}, {Rebbapragada}, {Reiley}, {Rigault}, {Rodriguez}, {van Roestel},
  {Rusholme}, {van Santen}, {Schulze}, {Shupe}, {Singer}, {Soumagnac}, {Stein},
  {Surace}, {Sollerman}, {Szkody}, {Taddia}, {Terek}, {Van Sistine}, {van
  Velzen}, {Vestrand}, {Walters}, {Ward}, {Ye}, {Yu}, {Yan}, \&
  {Zolkower}}]{Bellm2019}
{Bellm}, E.~C., {Kulkarni}, S.~R., {Graham}, M.~J., {et~al.}
  2019{\natexlab{a}}, 131, 018002, \dodoi{10.1088/1538-3873/aaecbe}

\bibitem[{{Bellm} {et~al.}(2019{\natexlab{b}}){Bellm}, {Kulkarni}, {Barlow},
  {Feindt}, {Graham}, {Goobar}, {Kupfer}, {Ngeow}, {Nugent}, {Ofek}, {Prince},
  {Riddle}, {Walters}, \& {Ye}}]{Bellm2019b}
{Bellm}, E.~C., {Kulkarni}, S.~R., {Barlow}, T., {et~al.} 2019{\natexlab{b}},
  \pasp, 131, 068003, \dodoi{10.1088/1538-3873/ab0c2a}

\bibitem[{{Bertini} {et~al.}(2017){Bertini}, {La Forgia}, {Tubiana},
  {G{\"u}ttler}, {Fulle}, {Moreno}, {Frattin}, {Kovacs}, {Pajola}, {Sierks},
  {Barbieri}, {Lamy}, {Rodrigo}, {Koschny}, {Rickman}, {Keller}, {Agarwal},
  {A'Hearn}, {Barucci}, {Bertaux}, {Bodewits}, {Cremonese}, {Da Deppo},
  {Davidsson}, {Debei}, {De Cecco}, {Drolshagen}, {Ferrari}, {Ferri},
  {Fornasier}, {Gicquel}, {Groussin}, {Gutierrez}, {Hasselmann}, {Hviid}, {Ip},
  {Jorda}, {Knollenberg}, {Kramm}, {K{\"u}hrt}, {K{\"u}ppers}, {Lara},
  {Lazzarin}, {Lin}, {Moreno}, {Lucchetti}, {Marzari}, {Massironi}, {Mottola},
  {Naletto}, {Oklay}, {Ott}, {Penasa}, {Thomas}, \& {Vincent}}]{Bertini2017}
{Bertini}, I., {La Forgia}, F., {Tubiana}, C., {et~al.} 2017, \mnras, 469,
  S404, \dodoi{10.1093/mnras/stx1850}

\bibitem[{{Bitsch} \& {Battistini}(2020)}]{Bitsch2020}
{Bitsch}, B., \& {Battistini}, C. 2020, \aap, 633, A10,
  \dodoi{10.1051/0004-6361/201936463}

\bibitem[{{Biver} {et~al.}(2018){Biver}, {Bockel{\'e}e-Morvan}, {Paubert},
  {Moreno}, {Crovisier}, {Boissier}, {Bertrand}, {Boussier}, {Kugel}, {McKay},
  {Dello Russo}, \& {DiSanti}}]{Biver2018}
{Biver}, N., {Bockel{\'e}e-Morvan}, D., {Paubert}, G., {et~al.} 2018, \aap,
  619, A127, \dodoi{10.1051/0004-6361/201833449}

\bibitem[{{Blagorodnova} {et~al.}(2018){Blagorodnova}, {Neill}, {Walters},
  {Kulkarni}, {Fremling}, {Ben-Ami}, {Dekany}, {Fucik}, {Konidaris}, {Nash},
  {Ngeow}, {Ofek}, {O' Sullivan}, {Quimby}, {Ritter}, \&
  {Vyhmeister}}]{Blagorodnova2018}
{Blagorodnova}, N., {Neill}, J.~D., {Walters}, R., {et~al.} 2018, \pasp, 130,
  035003, \dodoi{10.1088/1538-3873/aaa53f}

\bibitem[{{Bockel{\'e}e-Morvan} {et~al.}(2004){Bockel{\'e}e-Morvan},
  {Crovisier}, {Mumma}, \& {Weaver}}]{BockeleeMorvan2004}
{Bockel{\'e}e-Morvan}, D., {Crovisier}, J., {Mumma}, M.~J., \& {Weaver}, H.~A.
  2004, {The composition of cometary volatiles}, ed. M.~C. {Festou}, H.~U.
  {Keller}, \& H.~A. {Weaver}, 391

\bibitem[{{Bockel{\'e}e-Morvan} {et~al.}(2017){Bockel{\'e}e-Morvan}, {Rinaldi},
  {Erard}, {Leyrat}, {Capaccioni}, {Drossart}, {Filacchione}, {Migliorini},
  {Quirico}, {Mottola}, {Tozzi}, {Arnold}, {Biver}, {Combes}, {Crovisier},
  {Longobardo}, {Blecka}, \& {Capria}}]{BockeleeMorvan2017}
{Bockel{\'e}e-Morvan}, D., {Rinaldi}, G., {Erard}, S., {et~al.} 2017, \mnras,
  469, S443, \dodoi{10.1093/mnras/stx1950}

\bibitem[{{Bodewits} {et~al.}(2020){Bodewits}, {Noonan}, {Feldman},
  {Bannister}, {Farnocchia}, {Harris}, {Li}, {Mandt}, {Parker}, \&
  {Xing}}]{Bodewits2020}
{Bodewits}, D., {Noonan}, J.~W., {Feldman}, P.~D., {et~al.} 2020, Nature
  Astronomy, \dodoi{10.1038/s41550-020-1095-2}

\bibitem[{{Boe} {et~al.}(2019{\natexlab{a}}){Boe}, {Jedicke}, {Wiegert},
  {Meech}, {Morbidelli}, {Weryk}, \& {Morenz}}]{Boe2019b}
{Boe}, B., {Jedicke}, R., {Wiegert}, P., {et~al.} 2019{\natexlab{a}}, in
  EPSC-DPS Joint Meeting 2019, Vol. 2019, EPSC--DPS2019--626

\bibitem[{{Boe} {et~al.}(2019{\natexlab{b}}){Boe}, {Jedicke}, {Meech},
  {Wiegert}, {Weryk}, {Chambers}, {Denneau}, {Kaiser}, {Kudritzki}, {Magnier},
  {Wainscoat}, \& {Waters}}]{Boe2019}
{Boe}, B., {Jedicke}, R., {Meech}, K.~J., {et~al.} 2019{\natexlab{b}}, \icarus,
  333, 252, \dodoi{10.1016/j.icarus.2019.05.034}

\bibitem[{{Bolin}(2019)}]{Bolin2020hst}
{Bolin}, B.~T. 2019, arXiv e-prints, arXiv:1912.07386.
\newblock \doarXiv{1912.07386}

\bibitem[{{Bolin} {et~al.}(2018){Bolin}, {Weaver}, {Fernandez}, {Lisse},
  {Huppenkothen}, {Jones}, {Juri{\'c}}, {Moeyens}, {Schambeau}, {Slater},
  {Ivezi{\'c}}, \& {Connolly}}]{Bolin2018}
{Bolin}, B.~T., {Weaver}, H.~A., {Fernandez}, Y.~R., {et~al.} 2018, \apjl, 852,
  L2, \dodoi{10.3847/2041-8213/aaa0c9}

\bibitem[{{Bus} \& {Binzel}(2002)}]{Bus2002}
{Bus}, S.~J., \& {Binzel}, R.~P. 2002, Icarus, 158, 146,
  \dodoi{10.1006/icar.2002.6856}

\bibitem[{{Chambers} {et~al.}(2016){Chambers}, {Magnier}, {Metcalfe},
  {Flewelling}, {Huber}, {Waters}, {Denneau}, {Draper}, {Farrow}, {Finkbeiner},
  {Holmberg}, {Koppenhoefer}, {Price}, {Saglia}, {Schlafly}, {Smartt},
  {Sweeney}, {Wainscoat}, {Burgett}, {Grav}, {Heasley}, {Hodapp}, {Jedicke},
  {Kaiser}, {Kudritzki}, {Luppino}, {Lupton}, {Monet}, {Morgan}, {Onaka},
  {Stubbs}, {Tonry}, {Banados}, {Bell}, {Bender}, {Bernard}, {Botticella},
  {Casertano}, {Chastel}, {Chen}, {Chen}, {Cole}, {Deacon}, {Frenk},
  {Fitzsimmons}, {Gezari}, {Goessl}, {Goggia}, {Goldman}, {Grebel}, {Hambly},
  {Hasinger}, {Heavens}, {Heckman}, {Henderson}, {Henning}, {Holman}, {Hopp},
  {Ip}, {Isani}, {Keyes}, {Koekemoer}, {Kotak}, {Long}, {Lucey}, {Liu},
  {Martin}, {McLean}, {Morganson}, {Murphy}, {Nieto-Santisteban}, {Norberg},
  {Peacock}, {Pier}, {Postman}, {Primak}, {Rae}, {Rest}, {Riess}, {Riffeser},
  {Rix}, {Roser}, {Schilbach}, {Schultz}, {Scolnic}, {Szalay}, {Seitz},
  {Shiao}, {Small}, {Smith}, {Soderblom}, {Taylor}, {Thakar}, {Thiel},
  {Thilker}, {Urata}, {Valenti}, {Walter}, {Watters}, {Werner}, {White},
  {Wood-Vasey}, \& {Wyse}}]{Chambers2016}
{Chambers}, K.~C., {Magnier}, E.~A., {Metcalfe}, N., {et~al.} 2016, ArXiv
  e-prints.
\newblock \doarXiv{1612.05560}

\bibitem[{{Chen} {et~al.}(2006){Chen}, {Sargent}, {Bohac}, {Kim},
  {Leibensperger}, {Jura}, {Najita}, {Forrest}, {Watson}, {Sloan}, \&
  {Keller}}]{Chen2006}
{Chen}, C.~H., {Sargent}, B.~A., {Bohac}, C., {et~al.} 2006, \apjs, 166, 351,
  \dodoi{10.1086/505751}

\bibitem[{{Cochran} \& {McKay}(2018)}]{Cochran2018}
{Cochran}, A.~L., \& {McKay}, A.~J. 2018, \apjl, 854, L10,
  \dodoi{10.3847/2041-8213/aaab57}

\bibitem[{{Copandean} {et~al.}(2019){Copandean}, {Vaduvescu}, \&
  {Gorgan}}]{Copandean2019}
{Copandean}, D., {Vaduvescu}, O., \& {Gorgan}, D. 2019, arXiv e-prints,
  arXiv:1901.10469.
\newblock \doarXiv{1901.10469}

\bibitem[{{Cordiner} {et~al.}(2020){Cordiner}, {Milam}, {Biver},
  {Bockel{\'e}e-Morvan}, {Roth}, {Bergin}, {Jehin}, {Remijan}, {Charnley},
  {Mumma}, {Boissier}, {Crovisier}, {Paganini}, {Kuan}, \&
  {Lis}}]{Cordiner2020}
{Cordiner}, M.~A., {Milam}, S.~N., {Biver}, N., {et~al.} 2020, Nature
  Astronomy, \dodoi{10.1038/s41550-020-1087-2}

\bibitem[{{de Le{\'o}n} {et~al.}(2019{\natexlab{a}}){de Le{\'o}n}, {Licandro},
  {Serra-Ricart}, {Cabrera-Lavers}, {Font Serra}, {Scarpa}, {de la Fuente
  Marcos}, \& {de la Fuente Marcos}}]{deLeon2020}
{de Le{\'o}n}, J., {Licandro}, J., {Serra-Ricart}, M., {et~al.}
  2019{\natexlab{a}}, Research Notes of the American Astronomical Society, 3,
  131, \dodoi{10.3847/2515-5172/ab449c}

\bibitem[{{de Le{\'o}n} {et~al.}(2019{\natexlab{b}}){de Le{\'o}n}, {Licandro},
  {Serra-Ricart}, {Cabrera-Lavers}, {Font Serra}, {Scarpa}, {de la Fuente
  Marcos}, \& {de la Fuente Marcos}}]{deLeon2019}
---. 2019{\natexlab{b}}, Research Notes of the American Astronomical Society,
  3, 131, \dodoi{10.3847/2515-5172/ab449c}

\bibitem[{{DeMeo} {et~al.}(2009){DeMeo}, {Binzel}, {Slivan}, \&
  {Bus}}]{DeMeo2009}
{DeMeo}, F.~E., {Binzel}, R.~P., {Slivan}, S.~M., \& {Bus}, S.~J. 2009,
  \icarus, 202, 160, \dodoi{10.1016/j.icarus.2009.02.005}

\bibitem[{{DeMeo} \& {Carry}(2013)}]{DeMeo2013}
{DeMeo}, F.~E., \& {Carry}, B. 2013, \icarus, 226, 723,
  \dodoi{10.1016/j.icarus.2013.06.027}

\bibitem[{{Dohnanyi}(1969)}]{Dohnanyi1969}
{Dohnanyi}, J.~S. 1969, \jgr, 74, 2531, \dodoi{10.1029/JB074i010p02531}

\bibitem[{{Engelhardt} {et~al.}(2017){Engelhardt}, {Jedicke}, {Vere{\v s}},
  {Fitzsimmons}, {Denneau}, {Beshore}, \& {Meinke}}]{Engelhardt2017}
{Engelhardt}, T., {Jedicke}, R., {Vere{\v s}}, P., {et~al.} 2017, \aj, 153,
  133, \dodoi{10.3847/1538-3881/aa5c8a}

\bibitem[{{Feldman} {et~al.}(2004){Feldman}, {Cochran}, \&
  {Combi}}]{Feldman2004}
{Feldman}, P.~D., {Cochran}, A.~L., \& {Combi}, M.~R. 2004, {Spectroscopic
  investigations of fragment species in the coma}, ed. M.~C. {Festou}, H.~U.
  {Keller}, \& H.~A. {Weaver}, 425

\bibitem[{{Fern{\'a}ndez} {et~al.}(2001){Fern{\'a}ndez}, {Jewitt}, \&
  {Sheppard}}]{Fernandez2001}
{Fern{\'a}ndez}, Y.~R., {Jewitt}, D.~C., \& {Sheppard}, S.~S. 2001, \apjl, 553,
  L197, \dodoi{10.1086/320689}

\bibitem[{{Fern{\'a}ndez} {et~al.}(2013){Fern{\'a}ndez}, {Kelley}, {Lamy},
  {Toth}, {Groussin}, {Lisse}, {A'Hearn}, {Bauer}, {Campins}, {Fitzsimmons},
  {Licandro}, {Lowry}, {Meech}, {Pittichov{\'a}}, {Reach}, {Snodgrass}, \&
  {Weaver}}]{Fernandez2013}
{Fern{\'a}ndez}, Y.~R., {Kelley}, M.~S., {Lamy}, P.~L., {et~al.} 2013, \icarus,
  226, 1138, \dodoi{10.1016/j.icarus.2013.07.021}

\bibitem[{{Fink} \& {Rubin}(2012)}]{Fink2012}
{Fink}, U., \& {Rubin}, M. 2012, \icarus, 221, 721,
  \dodoi{10.1016/j.icarus.2012.09.001}

\bibitem[{{Fitzsimmons} {et~al.}(2017){Fitzsimmons}, {Hyland}, {Jedicke},
  {Snodgrass}, \& {Yang}}]{Fitzsimmons2017}
{Fitzsimmons}, A., {Hyland}, M., {Jedicke}, R., {Snodgrass}, C., \& {Yang}, B.
  2017, Central Bureau Electronic Telegrams, 4450

\bibitem[{{Fitzsimmons} {et~al.}(2018){Fitzsimmons}, {Snodgrass}, {Rozitis},
  {Yang}, {Hyland}, {Seccull}, {Bannister}, {Fraser}, {Jedicke}, \&
  {Lacerda}}]{Fitzsimmons2018}
{Fitzsimmons}, A., {Snodgrass}, C., {Rozitis}, B., {et~al.} 2018, Nature
  Astronomy, 2, 133, \dodoi{10.1038/s41550-017-0361-4}

\bibitem[{{Fitzsimmons} {et~al.}(2019){Fitzsimmons}, {Hainaut}, {Meech},
  {Jehin}, {Moulane}, {Opitom}, {Yang}, {Keane}, {Kleyna}, {Micheli}, \&
  {Snodgrass}}]{Fitzsimmons2019}
{Fitzsimmons}, A., {Hainaut}, O., {Meech}, K.~J., {et~al.} 2019, \apjl, 885,
  L9, \dodoi{10.3847/2041-8213/ab49fc}

\bibitem[{{Flewelling} {et~al.}(2016){Flewelling}, {Magnier}, {Chambers},
  {Heasley}, {Holmberg}, {Huber}, {Sweeney}, {Waters}, {Chen}, {Farrow},
  {Hasinger}, {Henderson}, {Long}, {Metcalfe}, {Nieto-Santisteban}, {Norberg},
  {Saglia}, {Szalay}, {Rest}, {Thakar}, {Tonry}, {Valenti}, {Werner}, {White},
  {Denneau}, {Draper}, {Hodapp}, {Jedicke}, {Kaiser}, {Kudritzki}, {Price},
  {Wainscoat}, {Chastel}, {McClean}, {Postman}, \& {Shiao}}]{Flewelling2016}
{Flewelling}, H.~A., {Magnier}, E.~A., {Chambers}, K.~C., {et~al.} 2016, ArXiv
  e-prints.
\newblock \doarXiv{1612.05243}

\bibitem[{{Graham} {et~al.}(2019){Graham}, {Kulkarni}, {Bellm}, {Adams},
  {Barbarino}, {Blagorodnova}, {Bodewits}, {Bolin}, {Brady}, \&
  {Cenko}}]{Graham2019}
{Graham}, M.~J., {Kulkarni}, S.~R., {Bellm}, E.~C., {et~al.} 2019, 131, 078001,
  \dodoi{10.1088/1538-3873/ab006c}

\bibitem[{{Hanu{\v s}} {et~al.}(2018){Hanu{\v s}}, {Delbo}, {Al{\'{\i}}-Lagoa},
  {Bolin}, {Jedicke}, {{\v D}urech}, {Cibulkov{\'a}}, {Pravec}, {Ku{\v
  s}nir{\'a}k}, {Behrend}, {Marchis}, {Antonini}, {Arnold}, {Audejean},
  {Bachschmidt}, {Bernasconi}, {Brunetto}, {Casulli}, {Dymock}, {Esseiva},
  {Esteban}, {Gerteis}, {de Groot}, {Gully}, {Hamanowa}, {Hamanowa}, {Krafft},
  {Lehk{\'y}}, {Manzini}, {Michelet}, {Morelle}, {Oey}, {Pilcher}, {Reignier},
  {Roy}, {Salom}, \& {Warner}}]{Hanus2018}
{Hanu{\v s}}, J., {Delbo}, M., {Al{\'{\i}}-Lagoa}, V., {et~al.} 2018, \icarus,
  299, 84, \dodoi{10.1016/j.icarus.2017.07.007}

\bibitem[{{Harker} {et~al.}(2018){Harker}, {Woodward}, {Kelley}, \&
  {Wooden}}]{Harker2018}
{Harker}, D.~E., {Woodward}, C.~E., {Kelley}, M. S.~P., \& {Wooden}, D.~H.
  2018, \aj, 155, 199, \dodoi{10.3847/1538-3881/aab778}

\bibitem[{{Huehnerhoff} {et~al.}(2016){Huehnerhoff}, {Ketzeback}, {Bradley},
  {Dembicky}, {Doughty}, {Hawley}, {Johnson}, {Klaene}, {Leon}, {McMillan},
  {Owen}, {Sayres}, {Sheen}, \& {Shugart}}]{Huehnerhoff2016}
{Huehnerhoff}, J., {Ketzeback}, W., {Bradley}, A., {et~al.} 2016, in \procspie,
  Vol. 9908, Ground-based and Airborne Instrumentation for Astronomy VI, 99085H

\bibitem[{{Hughes}(1990)}]{Hughes1990}
{Hughes}, D.~W. 1990, \qjras, 31, 69

\bibitem[{{Hui} {et~al.}(2020){Hui}, {Ye}, {F{\"o}hring}, {Hung}, \&
  {Tholen}}]{Hui2020a}
{Hui}, M.-T., {Ye}, Q.-Z., {F{\"o}hring}, D., {Hung}, D., \& {Tholen}, D.~J.
  2020, arXiv e-prints, arXiv:2003.14064.
\newblock \doarXiv{2003.14064}

\bibitem[{{Ivezi{\'c}} {et~al.}(2001){Ivezi{\'c}}, {Tabachnik}, {Rafikov},
  {Lupton}, {Quinn}, {Hammergren}, {Eyer}, {Chu}, {Armstrong}, {Fan},
  {Finlator}, {Geballe}, {Gunn}, {Hennessy}, {Knapp}, {Leggett}, {Munn},
  {Pier}, {Rockosi}, {Schneider}, {Strauss}, {Yanny}, {Brinkmann}, {Csabai},
  {Hindsley}, {Kent}, {Lamb}, {Margon}, {McKay}, {Smith}, {Waddel}, {York}, \&
  {SDSS Collaboration}}]{Ivezic2001}
{Ivezi{\'c}}, {\v Z}., {Tabachnik}, S., {Rafikov}, R., {et~al.} 2001, \aj, 122,
  2749, \dodoi{10.1086/323452}

\bibitem[{{Ivezi{\'c}} {et~al.}(2002){Ivezi{\'c}}, {Lupton}, {Juri{\'c}},
  {Tabachnik}, {Quinn}, {Gunn}, {Knapp}, {Rockosi}, \&
  {Brinkmann}}]{Ivezic2002}
{Ivezi{\'c}}, {\v Z}., {Lupton}, R.~H., {Juri{\'c}}, M., {et~al.} 2002, \aj,
  124, 2943, \dodoi{10.1086/344077}

\bibitem[{{Jedicke} {et~al.}(2016){Jedicke}, {Bolin}, {Granvik}, \&
  {Beshore}}]{Jedicke2016}
{Jedicke}, R., {Bolin}, B., {Granvik}, M., \& {Beshore}, E. 2016, \icarus, 266,
  173, \dodoi{10.1016/j.icarus.2015.10.021}

\bibitem[{{Jewitt}(1991)}]{Jewitt1991}
{Jewitt}, D. 1991, in Astrophysics and Space Science Library, Vol. 167, IAU
  Colloq. 116: Comets in the post-Halley era, ed. R.~L. {Newburn}, Jr.,
  M.~{Neugebauer}, \& J.~{Rahe}, 19--65

\bibitem[{{Jewitt} {et~al.}(2015){Jewitt}, {Hsieh}, \& {Agarwal}}]{Jewitt2015a}
{Jewitt}, D., {Hsieh}, H., \& {Agarwal}, J. 2015, {The Active Asteroids}, ed.
  P.~{Michel}, F.~E. {DeMeo}, \& W.~F. {Bottke}, 221--241

\bibitem[{{Jewitt} {et~al.}(2020){Jewitt}, {Hui}, {Kim}, {Mutchler}, {Weaver},
  \& {Agarwal}}]{Jewitt2019hst}
{Jewitt}, D., {Hui}, M.-T., {Kim}, Y., {et~al.} 2020, \apjl, 888, L23,
  \dodoi{10.3847/2041-8213/ab621b}

\bibitem[{{Jewitt} {et~al.}(2017{\natexlab{a}}){Jewitt}, {Hui}, {Mutchler},
  {Weaver}, {Li}, \& {Agarwal}}]{Jewitt2017cc}
{Jewitt}, D., {Hui}, M.-T., {Mutchler}, M., {et~al.} 2017{\natexlab{a}}, \apjl,
  847, L19, \dodoi{10.3847/2041-8213/aa88b4}

\bibitem[{{Jewitt} \& {Luu}(2019)}]{Jewitt2019}
{Jewitt}, D., \& {Luu}, J. 2019, \apjl, 886, L29,
  \dodoi{10.3847/2041-8213/ab530b}

\bibitem[{{Jewitt} {et~al.}(2017{\natexlab{b}}){Jewitt}, {Luu}, {Rajagopal},
  {Kotulla}, {Ridgway}, {Liu}, \& {Augusteijn}}]{Jewitt2017a}
{Jewitt}, D., {Luu}, J., {Rajagopal}, J., {et~al.} 2017{\natexlab{b}}, ArXiv
  e-prints.
\newblock \doarXiv{1711.05687}

\bibitem[{{Jewitt} \& {Meech}(1986)}]{Jewitt1986}
{Jewitt}, D., \& {Meech}, K.~J. 1986, \apj, 310, 937, \dodoi{10.1086/164745}

\bibitem[{{Jewitt} {et~al.}(2016){Jewitt}, {Mutchler}, {Weaver}, {Hui},
  {Agarwal}, {Ishiguro}, {Kleyna}, {Li}, {Meech}, {Micheli}, {Wainscoat}, \&
  {Weryk}}]{Jewitt2016}
{Jewitt}, D., {Mutchler}, M., {Weaver}, H., {et~al.} 2016, \apjl, 829, L8,
  \dodoi{10.3847/2041-8205/829/1/L8}

\bibitem[{{Jordi} {et~al.}(2006){Jordi}, {Grebel}, \& {Ammon}}]{Jordi2006}
{Jordi}, K., {Grebel}, E.~K., \& {Ammon}, K. 2006, \aap, 460, 339,
  \dodoi{10.1051/0004-6361:20066082}

\bibitem[{{Juri{\'c}} {et~al.}(2002){Juri{\'c}}, {Ivezi{\'c}}, {Lupton},
  {Quinn}, {Tabachnik}, {Fan}, {Gunn}, {Hennessy}, {Knapp}, {Munn}, {Pier},
  {Rockosi}, {Schneider}, {Brinkmann}, {Csabai}, \& {Fukugita}}]{Juric2002}
{Juri{\'c}}, M., {Ivezi{\'c}}, {\v Z}., {Lupton}, R.~H., {et~al.} 2002, \aj,
  124, 1776, \dodoi{10.1086/341950}

\bibitem[{{Kareta} {et~al.}(2019){Kareta}, {Andrews}, {Noonan}, {Harris},
  {Smith}, {O'Brien}, {Sharkey}, {Reddy}, {Springmann}, {Lejoly}, {Lunar},
  {Laboratory}, {:}, \& {Observatory}}]{Kareta2019}
{Kareta}, T., {Andrews}, J., {Noonan}, J.~W., {et~al.} 2019, arXiv e-prints,
  arXiv:1910.03222.
\newblock \doarXiv{1910.03222}

\bibitem[{{Kasliwal} {et~al.}(2019){Kasliwal}, {Cannella}, {Bagdasaryan},
  {Hung}, {Feindt}, {Singer}, {Coughlin}, {Fremling}, {Walters}, {Duev},
  {Itoh}, \& {Quimby}}]{Kasliwal2019}
{Kasliwal}, M.~M., {Cannella}, C., {Bagdasaryan}, A., {et~al.} 2019, \pasp,
  131, 038003, \dodoi{10.1088/1538-3873/aafbc2}

\bibitem[{{Keller} {et~al.}(2017){Keller}, {Mottola}, {Hviid}, {Agarwal},
  {K{\"u}hrt}, {Skorov}, {Otto}, {Vincent}, {Oklay}, {Schr{\"o}der},
  {Davidsson}, {Pajola}, {Shi}, {Bodewits}, {Toth}, {Preusker}, {Scholten},
  {Sierks}, {Barbieri}, {Lamy}, {Rodrigo}, {Koschny}, {Rickman}, {A'Hearn},
  {Barucci}, {Bertaux}, {Bertini}, {Cremonese}, {Da Deppo}, {Debei}, {De
  Cecco}, {Deller}, {Fornasier}, {Fulle}, {Groussin}, {Guti{\'e}rrez},
  {G{\"u}ttler}, {Hofmann}, {Ip}, {Jorda}, {Knollenberg}, {Kramm},
  {K{\"u}ppers}, {Lara}, {Lazzarin}, {Lopez-Moreno}, {Marzari}, {Naletto},
  {Tubiana}, \& {Thomas}}]{Keller2017}
{Keller}, H.~U., {Mottola}, S., {Hviid}, S.~F., {et~al.} 2017, \mnras, 469,
  S357, \dodoi{10.1093/mnras/stx1726}

\bibitem[{{Kelley} {et~al.}(2013){Kelley}, {Fern{\'a}ndez}, {Licandro},
  {Lisse}, {Reach}, {A'Hearn}, {Bauer}, {Campins}, {Fitzsimmons}, {Groussin},
  {Lamy}, {Lowry}, {Meech}, {Pittichov{\'a}}, {Snodgrass}, {Toth}, \&
  {Weaver}}]{Kelley2013}
{Kelley}, M.~S., {Fern{\'a}ndez}, Y.~R., {Licandro}, J., {et~al.} 2013,
  \icarus, 225, 475, \dodoi{10.1016/j.icarus.2013.04.012}

\bibitem[{{Kim} {et~al.}(2020){Kim}, {Jewitt}, {Mutchler}, {Agarwal}, {Hui}, \&
  {Weaver}}]{Kim2020}
{Kim}, Y., {Jewitt}, D., {Mutchler}, M., {et~al.} 2020, arXiv e-prints,
  arXiv:2005.02468.
\newblock \doarXiv{2005.02468}

\bibitem[{{Kinoshita} {et~al.}(2005){Kinoshita}, {Chen}, {Lin}, {Lin}, {Huang},
  {Chang}, \& {Chen}}]{Kinoshita2005}
{Kinoshita}, D., {Chen}, C.-W., {Lin}, H.-C., {et~al.} 2005, \cjaa, 5, 315,
  \dodoi{10.1088/1009-9271/5/3/011}

\bibitem[{{Kolokolova} {et~al.}(2004){Kolokolova}, {Hanner},
  {Levasseur-Regourd}, \& {Gustafson}}]{Kolokolova2004}
{Kolokolova}, L., {Hanner}, M.~S., {Levasseur-Regourd}, A.-C., \& {Gustafson},
  B.~{\AA}.~S. 2004, {Physical properties of cometary dust from light
  scattering and thermal emission}, ed. G.~W. {Kronk}, 577--604

\bibitem[{{Larkin} {et~al.}(2006){Larkin}, {Barczys}, {Krabbe}, {Adkins},
  {Aliado}, {Amico}, {Brims}, {Campbell}, {Canfield}, {Gasaway}, {Honey},
  {Iserlohe}, {Johnson}, {Kress}, {LaFreniere}, {Lyke}, {Magnone}, {Magnone},
  {McElwain}, {Moon}, {Quirrenbach}, {Skulason}, {Song}, {Spencer}, {Weiss}, \&
  {Wright}}]{Larkin2006}
{Larkin}, J., {Barczys}, M., {Krabbe}, A., {et~al.} 2006, in \procspie, Vol.
  6269, Society of Photo-Optical Instrumentation Engineers (SPIE) Conference
  Series, 62691A

\bibitem[{{Li} {et~al.}(2013){Li}, {Besse}, {A'Hearn}, {Belton}, {Bodewits},
  {Farnham}, {Klaasen}, {Lisse}, {Meech}, {Sunshine}, \& {Thomas}}]{Li2013}
{Li}, J.-Y., {Besse}, S., {A'Hearn}, M.~F., {et~al.} 2013, \icarus, 222, 559,
  \dodoi{10.1016/j.icarus.2012.11.001}

\bibitem[{{Li} {et~al.}(2016){Li}, {Samarasinha}, {Kelley}, {Farnham},
  {Bodewits}, {Lisse}, {Mutchler}, {A'Hearn}, \& {Delamere}}]{Li2016}
{Li}, J.-Y., {Samarasinha}, N.~H., {Kelley}, M. S.~P., {et~al.} 2016, \apjl,
  817, L23, \dodoi{10.3847/2041-8205/817/2/L23}

\bibitem[{{Lis} {et~al.}(2019){Lis}, {Bockel{\'e}e-Morvan}, {G{\"u}sten},
  {Biver}, {Stutzki}, {Delorme}, {Dur{\'a}n}, {Wiesemeyer}, \&
  {Okada}}]{Lis2019}
{Lis}, D.~C., {Bockel{\'e}e-Morvan}, D., {G{\"u}sten}, R., {et~al.} 2019, \aap,
  625, L5, \dodoi{10.1051/0004-6361/201935554}

\bibitem[{{Lisse} {et~al.}(2017){Lisse}, {Sitko}, {Marengo}, {Vervack},
  {Fernandez}, {Mittal}, \& {Chen}}]{Lisse2017}
{Lisse}, C.~M., {Sitko}, M.~L., {Marengo}, M., {et~al.} 2017, \aj, 154, 182,
  \dodoi{10.3847/1538-3881/aa855e}

\bibitem[{{Lisse} {et~al.}(2019){Lisse}, {Young}, {Cruikshank}, {Sandford}, \&
  {Schmitt}}]{Lisse2019}
{Lisse}, C.~M., {Young}, L.~A., {Cruikshank}, D., {Sandford}, S., \& {Schmitt},
  B. 2019, Submitted to Icarus

\bibitem[{{Lisse} {et~al.}(2009){Lisse}, {Fernandez}, {Reach}, {Bauer},
  {A'Hearn}, {Farnham}, {Groussin}, {Belton}, {Meech}, \&
  {Snodgrass}}]{Lisse2009}
{Lisse}, C.~M., {Fernandez}, Y.~R., {Reach}, W.~T., {et~al.} 2009, \pasp, 121,
  968, \dodoi{10.1086/605546}

\bibitem[{{Lisse} {et~al.}(2012){Lisse}, {Wyatt}, {Chen}, {Morlok}, {Watson},
  {Manoj}, {Sheehan}, {Currie}, {Thebault}, \& {Sitko}}]{Lisse2012}
{Lisse}, C.~M., {Wyatt}, M.~C., {Chen}, C.~H., {et~al.} 2012, \apj, 747, 93,
  \dodoi{10.1088/0004-637X/747/2/93}

\bibitem[{{MacGregor} {et~al.}(2019){MacGregor}, {Weinberger}, {Nesvold},
  {Hughes}, {Wilner}, {Currie}, {Debes}, {Donaldson}, {Redfield}, {Roberge}, \&
  {Schneider}}]{MacGregor2019}
{MacGregor}, M.~A., {Weinberger}, A.~J., {Nesvold}, E.~R., {et~al.} 2019,
  \apjl, 877, L32, \dodoi{10.3847/2041-8213/ab21c2}

\bibitem[{{Marchis} {et~al.}(2006){Marchis}, {Kaasalainen}, {Hom}, {Berthier},
  {Enriquez}, {Hestroffer}, {Le Mignant}, \& {de Pater}}]{Marchis2006}
{Marchis}, F., {Kaasalainen}, M., {Hom}, E.~F.~Y., {et~al.} 2006, \icarus, 185,
  39, \dodoi{10.1016/j.icarus.2006.06.001}

\bibitem[{{Masci} {et~al.}(2019){Masci}, {Laher}, {Rusholme}, {Shupe}, {Groom},
  {Surace}, {Jackson}, {Monkewitz}, {Beck}, {Flynn}, {Terek}, {Landry},
  {Hacopians}, {Desai}, {Howell}, {Brooke}, {Imel}, {Wachter}, {Ye}, {Lin},
  {Cenko}, {Cunningham}, {Rebbapragada}, {Bue}, {Miller}, {Mahabal}, {Bellm},
  {Patterson}, {Juri{\'c}}, {Golkhou}, {Ofek}, {Walters}, {Graham}, {Kasliwal},
  {Dekany}, {Kupfer}, {Burdge}, {Cannella}, {Barlow}, {Van Sistine}, {Giomi},
  {Fremling}, {Blagorodnova}, {Levitan}, {Riddle}, {Smith}, {Helou}, {Prince},
  \& {Kulkarni}}]{Masci2019}
{Masci}, F.~J., {Laher}, R.~R., {Rusholme}, B., {et~al.} 2019, \pasp, 131,
  018003, \dodoi{10.1088/1538-3873/aae8ac}

\bibitem[{{McKay} {et~al.}(2019){McKay}, {Cochran}, {Dello Russo}, \&
  {DiSanti}}]{McKay2019}
{McKay}, A.~J., {Cochran}, A.~L., {Dello Russo}, N., \& {DiSanti}, M. 2019,
  arXiv e-prints, arXiv:1910.12785.
\newblock \doarXiv{1910.12785}

\bibitem[{{Meech}(2017)}]{Meech2017b}
{Meech}, K.~J. 2017, Philosophical Transactions of the Royal Society of London
  Series A, 375, 20160247, \dodoi{10.1098/rsta.2016.0247}

\bibitem[{{Meech} \& {Svoren}(2004)}]{Meech2004}
{Meech}, K.~J., \& {Svoren}, J. 2004, {Using cometary activity to trace the
  physical and chemical evolution of cometary nuclei}, ed. M.~C. {Festou},
  H.~U. {Keller}, \& H.~A. {Weaver}, 317

\bibitem[{Meech {et~al.}(2017)Meech, Weryk, Micheli, Kleyna, Hainaut, Jedicke,
  Wainscoat, Chambers, Keane, Petric, Denneau, Magnier, Berger, Huber,
  Flewelling, Waters, Schunova-Lilly, \& Chastel}]{Meech2017}
Meech, K.~J., Weryk, R., Micheli, M., {et~al.} 2017, Nature, EP

\bibitem[{{Meng} {et~al.}(2014){Meng}, {Su}, {Rieke}, {Stevenson}, {Plavchan},
  {Rujopakarn}, {Lisse}, {Poshyachinda}, \& {Reichart}}]{Meng2014}
{Meng}, H.~Y.~A., {Su}, K.~Y.~L., {Rieke}, G.~H., {et~al.} 2014, Science, 345,
  1032, \dodoi{10.1126/science.1255153}

\bibitem[{{Micheli} {et~al.}(2018){Micheli}, {Farnocchia}, {Meech}, {Buie},
  {Hainaut}, {Prialnik}, {Sch{\"o}rghofer}, {Weaver}, {Chodas}, {Kleyna},
  {Weryk}, {Wainscoat}, {Ebeling}, {Keane}, {Chambers}, {Koschny}, \&
  {Petropoulos}}]{Micheli2018}
{Micheli}, M., {Farnocchia}, D., {Meech}, K.~J., {et~al.} 2018, \nat, 559, 223,
  \dodoi{10.1038/s41586-018-0254-4}

\bibitem[{{Morbidelli} \& {Nesvorny}(2019)}]{Morbidelli2019}
{Morbidelli}, A., \& {Nesvorny}, D. 2019, arXiv e-prints, arXiv:1904.02980.
\newblock \doarXiv{1904.02980}

\bibitem[{{Moreno} {et~al.}(2017){Moreno}, {Pozuelos}, {Novakovi{\'c}},
  {Licandro}, {Cabrera-Lavers}, {Bolin}, {Jedicke}, {Gladman}, {Bannister},
  {Gwyn}, {Vere{\v s}}, {Chambers}, {Chastel}, {Denneau}, {Flewelling},
  {Huber}, {Schunov{\'a}-Lilly}, {Magnier}, {Wainscoat}, {Waters}, {Weryk},
  {Farnocchia}, \& {Micheli}}]{Moreno2017}
{Moreno}, F., {Pozuelos}, F.~J., {Novakovi{\'c}}, B., {et~al.} 2017, \apjl,
  837, L3, \dodoi{10.3847/2041-8213/aa6036}

\bibitem[{Ofek(2012)}]{Ofek2012}
Ofek, E.~O. 2012, The Astrophysical Journal, 749, 10

\bibitem[{{Opitom} {et~al.}(2019){Opitom}, {Fitzsimmons}, {Jehin}, {Moulane},
  {Hainaut}, {Meech}, {Yang}, {Snodgrass}, {Micheli}, {Keane}, {Benkhaldoun},
  \& {Kleyna}}]{Opitom2019}
{Opitom}, C., {Fitzsimmons}, A., {Jehin}, E., {et~al.} 2019, arXiv e-prints,
  arXiv:1910.09078.
\newblock \doarXiv{1910.09078}

\bibitem[{{Paganini} {et~al.}(2014){Paganini}, {Mumma}, {Villanueva}, {Keane},
  {Blake}, {Bonev}, {DiSanti}, {Gibb}, \& {Meech}}]{Paganini2014}
{Paganini}, L., {Mumma}, M.~J., {Villanueva}, G.~L., {et~al.} 2014, \apj, 791,
  122, \dodoi{10.1088/0004-637X/791/2/122}

\bibitem[{{P{\"a}tzold} {et~al.}(2016){P{\"a}tzold}, {Andert}, {Hahn}, {Asmar},
  {Barriot}, {Bird}, {H{\"a}usler}, {Peter}, {Tellmann}, {Gr{\"u}n},
  {Weissman}, {Sierks}, {Jorda}, {Gaskell}, {Preusker}, \&
  {Scholten}}]{Patzold2016}
{P{\"a}tzold}, M., {Andert}, T., {Hahn}, M., {et~al.} 2016, \nat, 530, 63,
  \dodoi{10.1038/nature16535}

\bibitem[{{Protopapa} {et~al.}(2014){Protopapa}, {Sunshine}, {Feaga}, {Kelley},
  {A'Hearn}, {Farnham}, {Groussin}, {Besse}, {Merlin}, \& {Li}}]{Protopapa2014}
{Protopapa}, S., {Sunshine}, J.~M., {Feaga}, L.~M., {et~al.} 2014, \icarus,
  238, 191, \dodoi{10.1016/j.icarus.2014.04.008}

\bibitem[{{Rappaport} {et~al.}(2018){Rappaport}, {Vanderburg}, {Jacobs},
  {LaCourse}, {Jenkins}, {Kraus}, {Rizzuto}, {Latham}, {Bieryla}, {Lazarevic},
  \& {Schmitt}}]{Rappaport2018}
{Rappaport}, S., {Vanderburg}, A., {Jacobs}, T., {et~al.} 2018, \mnras, 474,
  1453, \dodoi{10.1093/mnras/stx2735}

\bibitem[{{Raymond} {et~al.}(2018{\natexlab{a}}){Raymond}, {Armitage}, \&
  {Veras}}]{Raymond2018b}
{Raymond}, S.~N., {Armitage}, P.~J., \& {Veras}, D. 2018{\natexlab{a}}, \apjl,
  856, L7, \dodoi{10.3847/2041-8213/aab4f6}

\bibitem[{{Raymond} {et~al.}(2018{\natexlab{b}}){Raymond}, {Armitage}, \&
  {Veras}}]{Raymond2018bc}
---. 2018{\natexlab{b}}, \apjl, 856, L7, \dodoi{10.3847/2041-8213/aab4f6}

\bibitem[{{Raymond} {et~al.}(2018{\natexlab{c}}){Raymond}, {Armitage}, {Veras},
  {Quintana}, \& {Barclay}}]{Raymond2018a}
{Raymond}, S.~N., {Armitage}, P.~J., {Veras}, D., {Quintana}, E.~V., \&
  {Barclay}, T. 2018{\natexlab{c}}, \mnras, 476, 3031,
  \dodoi{10.1093/mnras/sty468}

\bibitem[{{Rayner} {et~al.}(2003){Rayner}, {Toomey}, {Onaka}, {Denault},
  {Stahlberger}, {Vacca}, {Cushing}, \& {Wang}}]{Rayner2003}
{Rayner}, J.~T., {Toomey}, D.~W., {Onaka}, P.~M., {et~al.} 2003, \pasp, 115,
  362, \dodoi{10.1086/367745}

\bibitem[{{Rigault} {et~al.}(2019){Rigault}, {Neill}, {Blagorodnova}, {Dugas},
  {Feeney}, {Walters}, {Brinnel}, {Copin}, {Fremling}, {Nordin}, \&
  {Sollerman}}]{Rigault2019}
{Rigault}, M., {Neill}, J.~D., {Blagorodnova}, N., {et~al.} 2019, \aap, 627,
  A115, \dodoi{10.1051/0004-6361/201935344}

\bibitem[{{Schwamb} {et~al.}(2019){Schwamb}, {Fraser}, {Bannister}, {Marsset},
  {Pike}, {Kavelaars}, {Benecchi}, {Lehner}, {Wang}, {Thirouin}, {Delsanti},
  {Peixinho}, {Volk}, {Alexandersen}, {Chen}, {Gladman}, {Gwyn}, \&
  {Petit}}]{Schwamb2019}
{Schwamb}, M.~E., {Fraser}, W.~C., {Bannister}, M.~T., {et~al.} 2019, \apjs,
  243, 12, \dodoi{10.3847/1538-4365/ab2194}

\bibitem[{{Singer} {et~al.}(2019){Singer}, {McKinnon}, {Gladman},
  {Greenstreet}, {Bierhaus}, {Stern}, {Parker}, {Robbins}, {Schenk}, {Grundy},
  {Bray}, {Beyer}, {Binzel}, {Weaver}, {Young}, {Spencer}, {Kavelaars},
  {Moore}, {Zangari}, {Olkin}, {Lauer}, {Lisse}, {Ennico}, {New Horizons
  Geology}, Team, {New Horizons Surface Composition Science Theme Team}, \&
  {New Horizons Ralph and LORRI Teams}}]{Singer2019}
{Singer}, K.~N., {McKinnon}, W.~B., {Gladman}, B., {et~al.} 2019, Science, 363,
  955, \dodoi{10.1126/science.aap8628}

\bibitem[{{Skrutskie} {et~al.}(2006){Skrutskie}, {Cutri}, {Stiening},
  {Weinberg}, {Schneider}, {Carpenter}, {Beichman}, {Capps}, {Chester},
  {Elias}, {Huchra}, {Liebert}, {Lonsdale}, {Monet}, {Price}, {Seitzer},
  {Jarrett}, {Kirkpatrick}, {Gizis}, {Howard}, {Evans}, {Fowler}, {Fullmer},
  {Hurt}, {Light}, {Kopan}, {Marsh}, {McCallon}, {Tam}, {Van Dyk}, \&
  {Wheelock}}]{Skrutskie2006}
{Skrutskie}, M.~F., {Cutri}, R.~M., {Stiening}, R., {et~al.} 2006, \aj, 131,
  1163, \dodoi{10.1086/498708}

\bibitem[{{Smith} \& {Nelson}(1969)}]{Smith_Nelson1969}
{Smith}, C.~E., \& {Nelson}, B. 1969, \pasp, 81, 74, \dodoi{10.1086/128742}

\bibitem[{{Snodgrass} {et~al.}(2017){Snodgrass}, {Agarwal}, {Combi},
  {Fitzsimmons}, {Guilbert-Lepoutre}, {Hsieh}, {Hui}, {Jehin}, {Kelley},
  {Knight}, {Opitom}, {Orosei}, {de Val-Borro}, \& {Yang}}]{Snodgrass2017}
{Snodgrass}, C., {Agarwal}, J., {Combi}, M., {et~al.} 2017, \aapr, 25, 5,
  \dodoi{10.1007/s00159-017-0104-7}

\bibitem[{{Solontoi} {et~al.}(2012){Solontoi}, {Ivezi{\'c}}, {Juri{\'c}},
  {Becker}, {Jones}, {West}, {Kent}, {Lupton}, {Claire}, {Knapp}, {Quinn},
  {Gunn}, \& {Schneider}}]{Solontoi2012}
{Solontoi}, M., {Ivezi{\'c}}, {\v Z}., {Juri{\'c}}, M., {et~al.} 2012, \icarus,
  218, 571, \dodoi{10.1016/j.icarus.2011.10.008}

\bibitem[{Steele {et~al.}(2004)Steele, Smith, Rees, Baker, Bates, Bode, Bowman,
  Carter, Etherton, Ford, Fraser, Gomboc, Lett, Mansfield, Marchant,
  Medrano-Cerda, Mottram, Raback, Scott, Tomlinson, \& Zamanov}]{Steele2004}
Steele, I.~A., Smith, R.~J., Rees, P. C.~T., {et~al.} 2004, in SPIE
  Astronomical Telescopes + Instrumentation

\bibitem[{{Su} {et~al.}(2019){Su}, {Jackson}, {G{\'a}sp{\'a}r}, {Rieke},
  {Dong}, {Olofsson}, {Kennedy}, {Leinhardt}, {Malhotra}, {Hammer}, {Meng},
  {Rujopakarn}, {Rodriguez}, {Pepper}, {Reichart}, {James}, \&
  {Stassun}}]{Su2019}
{Su}, K. Y.~L., {Jackson}, A.~P., {G{\'a}sp{\'a}r}, A., {et~al.} 2019, \aj,
  157, 202, \dodoi{10.3847/1538-3881/ab1260}

\bibitem[{{Tonry} {et~al.}(2012){Tonry}, {Stubbs}, {Lykke}, {Doherty},
  {Shivvers}, {Burgett}, {Chambers}, {Hodapp}, {Kaiser}, {Kudritzki},
  {Magnier}, {Morgan}, {Price}, \& {Wainscoat}}]{Tonry2012}
{Tonry}, J.~L., {Stubbs}, C.~W., {Lykke}, K.~R., {et~al.} 2012, \apj, 750, 99,
  \dodoi{10.1088/0004-637X/750/2/99}

\bibitem[{{Trilling} {et~al.}(2017){Trilling}, {Robinson}, {Roegge}, {Chand
  ler}, {Smith}, {Loeffler}, {Trujillo}, {Navarro-Meza}, \&
  {Glaspie}}]{Trilling2018a}
{Trilling}, D.~E., {Robinson}, T., {Roegge}, A., {et~al.} 2017, \apjl, 850,
  L38, \dodoi{10.3847/2041-8213/aa9989}

\bibitem[{{Trilling} {et~al.}(2018){Trilling}, {Mommert}, {Hora}, {Farnocchia},
  {Chodas}, {Giorgini}, {Smith}, {Carey}, {Lisse}, {Werner}, {McNeill},
  {Chesley}, {Emery}, {Fazio}, {Fernandez}, {Harris}, {Marengo}, {Mueller},
  {Roegge}, {Smith}, {Weaver}, {Meech}, \& {Micheli}}]{Trilling2018}
{Trilling}, D.~E., {Mommert}, M., {Hora}, J.~L., {et~al.} 2018, \aj, 156, 261,
  \dodoi{10.3847/1538-3881/aae88f}

\bibitem[{{Vincent} {et~al.}(2003){Vincent}, {Morse}, {Beland}, {Hearty},
  {Bally}, {Ellingson}, {Wilkinson}, {Hartigan}, {Holtzman}, \&
  {Barentine}}]{Vincent2003}
{Vincent}, M.~B., {Morse}, J.~A., {Beland}, S., {et~al.} 2003, in Society of
  Photo-Optical Instrumentation Engineers (SPIE) Conference Series, Vol. 4841,
  \procspie, ed. M.~{Iye} \& A.~F.~M. {Moorwood}, 367--375

\bibitem[{{Vokrouhlick{\'y}} {et~al.}(2017){Vokrouhlick{\'y}}, {Pravec},
  {Durech}, {Bolin}, {Jedicke}, {Ku{\v s}nir{\'a}k}, {Gal{\'a}d}, {Hornoch},
  {Kryszczy{\'n}ska}, {Colas}, {Moskovitz}, {Thirouin}, \&
  {Nesvorn{\'y}}}]{Vokrouhlicky2017a}
{Vokrouhlick{\'y}}, D., {Pravec}, P., {Durech}, J., {et~al.} 2017, \aap, 598,
  A91, \dodoi{10.1051/0004-6361/201629670}

\bibitem[{{Williams}(2017)}]{Williams2017}
{Williams}, G.~V. 2017, Minor Planet Electronic Circulars

\bibitem[{{Williams}(2019{\natexlab{a}})}]{Williams2019a}
---. 2019{\natexlab{a}}, Minor Planet Electronic Circulars

\bibitem[{{Williams}(2019{\natexlab{b}})}]{Williams2019b}
---. 2019{\natexlab{b}}, Minor Planet Electronic Circulars

\bibitem[{{Williams}(2019{\natexlab{c}})}]{Williams2019bb}
---. 2019{\natexlab{c}}, Minor Planet Electronic Circulars

\bibitem[{{Womack} {et~al.}(2017){Womack}, {Sarid}, \&
  {Wierzchos}}]{Womack2017}
{Womack}, M., {Sarid}, G., \& {Wierzchos}, K. 2017, \pasp, 129, 031001,
  \dodoi{10.1088/1538-3873/129/973/031001}

\bibitem[{{Yang} {et~al.}(2009){Yang}, {Jewitt}, \& {Bus}}]{Yang2009}
{Yang}, B., {Jewitt}, D., \& {Bus}, S.~J. 2009, \aj, 137, 4538,
  \dodoi{10.1088/0004-6256/137/5/4538}

\bibitem[{{Yang} {et~al.}(2019){Yang}, {Keane}, {Kelley}, {Protopapa}, \&
  {Meech}}]{Yang2019}
{Yang}, B., {Keane}, J., {Kelley}, M., {Protopapa}, S., \& {Meech}, K.~J. 2019,
  Central Bureau Electronic Telegrams, 4672

\bibitem[{{Ye} {et~al.}(2019){Ye}, {Kelley}, {Bolin}, {Bodewits}, {Farnocchia},
  {Masci}, {Meech}, {Micheli}, {Weryk}, {Bellm}, {Christensen}, {Dekany},
  {Delacroix}, {Graham}, {Kulkarni}, {Laher}, {Rusholme}, \& {Smith}}]{Ye2019b}
{Ye}, Q., {Kelley}, M. S.~P., {Bolin}, B.~T., {et~al.} 2019, arXiv e-prints,
  arXiv:1911.05902.
\newblock \doarXiv{1911.05902}

\bibitem[{{Zhang} \& {Lin}(2020)}]{Zhang2020}
{Zhang}, Y., \& {Lin}, D. N.~C. 2020, Nature Astronomy,
  \dodoi{10.1038/s41550-020-1065-8}

\end{thebibliography}

\begin{longtable}{|c|c|c|c|c|c|c|c|c|c|c|}
\caption{Summary of comet 2I photometry.\label{tab:phot}}\\
\hline
Date$^1$ & Telescope$^2$ & $r_h^3$&$\Delta^4$&$\alpha^5$&$\nu^6$&filter$^7$& mag$^8$ & $\sigma_\mathrm{mag}^9$ & $\theta_s^{10}$ & $\chi_{am}^{11}$ \\
UTC&&(au)&(au)&($^{\circ}$)&($^{\circ}$)&&&&($\arcsec$)&\\
\hline
\endfirsthead
\multicolumn{4}{c}%
{\tablename\ \thetable\ -- \textit{Continued from previous page}} \\
\hline
Date$^1$ & Telescope$^2$ & $r_h^3$&$\Delta^4$&$\alpha^5$&$\nu^6$&filter$^7$& mag$^8$ & $\sigma_\mathrm{mag}^9$ & $\theta_s^{10}$ & $\chi_{am}^{11}$ \\
UTC&&(au)&(au)&($^{\circ}$)&($^{\circ}$)&&&&($\arcsec$)&\\
\hline
\endhead
\hline \multicolumn{4}{r}{\textit{Continued on next page}} \\
\endfoot
\hline
\endlastfoot
2019 Mar 17 & ZTF & 6.0 & 6.1 & 9.3 & 277.6 & r & 20.71 & 0.37&2.37&1.32\\
2019 Mar 18 & ZTF & 6.0 & 6.1 & 9.3 & 277.7 & r & 21.01& 0.37&2.18&1.27\\
2019 May 02 & ZTF & 5.2 & 5.8 & 8.2 & 281.8 & r & 20.30& 0.18&2.11&1.90\\
2019 May 05 & ZTF & 5.1 & 5.8 & 8.0 & 282.0 & r & 20.66& 0.31&2.53&2.01\\
2019 Sep 10 & SEDM & 2.8 & 3.48 & 13.9 & 308.6 & r & 17.91& 0.05&2.11&1.89\\
2019 Sep 11 & SEDM & 2.8 & 3.45 & 14.1 & 309.2 & r & 17.71& 0.04&1.74&1.68\\
2019 Sep 11 & ZTF & 2.8 & 3.45 & 14.1 & 309.2 & g & 18.43& 0.06&1.92&1.40\\
2019 Sep 12 & ARC 3.5 m & 2.78 & 3.43 & 14.3 & 309.6 & g & 18.29& 0.04&0.66&2.34\\
2019 Sep 12 & ARC 3.5 m & 2.78 & 3.43 & 14.3 & 309.6 & r & 17.75& 0.04&0.62&2.15\\
2019 Sep 12 & ARC 3.5 m & 2.78 & 3.43 & 14.3 & 309.6 & i & 17.55& 0.01&0.58&2.32\\
2019 Sep 12 & ARC 3.5 m & 2.78 & 3.43 & 14.3 & 309.6 & z & 17.78& 0.03&0.56&2.21\\
2019 Sep 12 & Lulin& 2.78 & 3.43 & 14.3 & 309.6 & V & 18.01& 0.05&3.50&2.36\\
2019 Sep 12 & Lulin& 2.78 & 3.43 & 14.3 & 309.6 & B & 18.77& 0.1&3.50&2.36\\
2019 Sep 12 & Lulin & 2.78 & 3.43 & 14.3 & 309.6 & R & 17.47& 0.04&3.50&2.36\\
2019 Sep 12 & Lulin & 2.78 & 3.43 & 14.3 & 309.6 & I & 17.09& 0.04&3.50&2.36\\
2019 Sep 15 & Bisei & 2.73 & 3.37 & 14.9 & 310.7 & R & 17.41& 0.06&2.05&2.11\\
2019 Sep 18 & Liverpool & 2.69 & 3.30 & 15.5 & 311.9 & g & 18.23& 0.17&1.03&2.25\\
2019 Sep 19 & MLO 1.0-m & 2.68 & 3.27 & 15.7 & 312.3 & r & 17.86 & 0.02&3.53&2.08\\
2019 Sep 21 & ZTF & 2.65 & 3.23 & 16.0 & 313.1 & r & 17.82& 0.04&1.76&1.76\\
2019 Sep 22 & ZTF & 2.64 & 3.21 & 16.2 & 313.6 & g & 17.72 & 0.06&2.53&1.72\\
2019 Sep 22 & ZTF & 2.64 & 3.21 & 16.2 & 313.6 & r & 18.33& 0.13&2.16&2.21\\
2019 Sep 27 & ARC 3.5 m & 2.56 & 3.11 & 17.2 & 315.7 & R & 17.45& 0.04&1.87&2.25\\
2019 Sep 30 & MLO 1.0-m & 2.52 & 3.04 & 17.8 & 317.1 & r & 17.51& 0.02&3.49&2.32\\
2019 Oct 01 & ZTF & 2.51 & 3.02 & 18.0 & 317.5 & r & 17.36 & 0.03&2.20&2.35\\
2019 Oct 02 & ZTF & 2.50 & 2.99 & 18.2 & 318.0 & g & 17.83 & 0.04&2.24&2.20\\
2019 Oct 02 & ZTF & 2.50 & 2.99 & 18.2 & 318.0 & r & 17.25 & 0.04&1.94&1.80\\
2019 Oct 04 & MLO 1.0-m & 2.47 & 2.95 & 18.64 & 318.9 & V & 17.80& 0.02&3.75&1.99\\
2019 Oct 04 & MLO 1.0-m & 2.47 & 2.95 & 18.64 & 318.9 & r & 17.32& 0.01&3.62&1.95\\
2019 Oct 04 & Liverpool & 2.47 & 2.95 & 18.64 & 318.9 & g & 17.89& 0.02&1.12&1.97\\
2019 Oct 04 & Liverpool & 2.47 & 2.95 & 18.64 & 318.9 & r & 17.31& 0.01&1.17&1.94\\
2019 Oct 05 & ZTF & 2.46 & 2.94 & 18.85 & 319.4 & r & 17.19 & 0.04&1.97&2.02\\
2019 Oct 08 & MLO 1.0-m & 2.42 & 2.87 & 19.45 & 320.8 & V & 17.73& 0.03&2.82&2.17\\
2019 Oct 08 & MLO 1.0-m & 2.42 & 2.87 & 19.45 & 320.8 & r & 17.21& 0.01&2.95&2.22\\
2019 Oct 08 & Liverpool & 2.42 & 2.87 & 19.45 & 320.8 & g & 17.75& 0.03&1.03&2.01\\
2019 Oct 08 & Liverpool & 2.42 & 2.87 & 19.45 & 320.8 & r & 17.15& 0.01&1.07&1.99\\
2019 Oct 10 & Liverpool & 2.40 & 2.83 & 19.85 & 321.8 & g & 17.61& 0.01&0.99&1.98\\
2019 Oct 10 & Liverpool & 2.40 & 2.83 & 19.85 & 321.8 & r & 17.13& 0.01&1.08&1.93\\
2019 Oct 11 & ZTF & 2.38 & 2.81 & 20.1 & 322.4 & g & 17.71 & 0.04&2.99&2.03\\
2019 Oct 12 & ARC 3.5 m & 2.37 & 2.79 & 20.25 & 322.9 & B & 18.04& 0.04&0.88&1.56\\
2019 Oct 12 & ARC 3.5 m & 2.37 & 2.79 & 20.25 & 322.9 & g & 17.74& 0.05&0.73&1.71\\
2019 Oct 12 & ARC 3.5 m & 2.37 & 2.79 & 20.25 & 322.9 & r & 17.11& 0.05&0.78&2.02\\
2019 Oct 12 & ARC 3.5 m & 2.37 & 2.79 & 20.25 & 322.9 & i & 16.94& 0.02&0.82&2.12\\
2019 Oct 12 & ARC 3.5 m & 2.37 & 2.79 & 20.25 & 322.9 & z & 17.14& 0.05&0.75&1.62\\
2019 Oct 12 & MLO 1.0-m & 2.37 & 2.79 & 20.25 & 322.9 & V & 17.55& 0.04&2.78&1.74\\
2019 Oct 12 & MLO 1.0-m & 2.37 & 2.79 & 20.25 & 322.9 & r & 17.27& 0.04&2.56&1.83\\
2019 Oct 14 & Liverpool & 2.35 & 2.76 & 20.66 & 323.9 & r & 17.14& 0.04&1.10&1.9\\
2019 Oct 14 & ZTF & 2.35 & 2.76 & 20.66 & 323.9 & g & 17.74& 0.08&2.00&1.37\\
2019 Oct 14 & ZTF & 2.35 & 2.76 & 20.66 & 323.9 & r & 17.16& 0.04&2.01&1.77\\
2019 Oct 15 & ZTF & 2.34 & 2.74 & 20.86 & 324.4 & r & 17.18& 0.05&3.84&1.79\\
2019 Oct 17 & MLO 1.0-m & 2.32 & 2.70 & 21.26 & 325.5 & V & 17.46& 0.04&2.93&1.68\\
2019 Oct 17 & MLO 1.0-m & 2.32 & 2.70 & 21.26 & 325.5 & r & 17.19& 0.04&3.12&1.76\\
2019 Oct 21 & ARC 3.5 m & 2.28 & 2.62 & 22.04 & 327.7 & r & 16.99& 0.02&2.21&2.13\\
2019 Oct 29 & ZTF & 2.20 & 2.48 & 23.56 & 332.4 & r & 16.84 & 0.01&1.92&1.59\\
2019 Nov 03 & ARC 3.5 m & 2.16 & 2.40 & 24.46 & 335.5 & r & 16.76 & 0.02&0.96&2.58\\
2019 Nov 05 & ZTF & 2.14 & 2.37 & 24.79 & 336.7 & r & 16.61 & 0.03&1.74&1.68\\
2019 Nov 08 & ZTF & 2.12 & 2.32 & 25.29 & 338.7 & r & 16.76 & 0.03&1.92 &1.40\\
2019 Nov 12 & ZTF & 2.09 & 2.26 & 25.91 & 341.3 & r & 16.76 & 0.04&2.13 &2.20\\
2019 Nov 17 & ZTF & 2.06 & 2.20 & 26.62 & 344.7 & r & 16.88 & 0.04&2.67 &1.40\\
2019 Nov 27 & ZTF & 2.02 & 2.08 & 27.76 & 351.7 & g & 17.29 & 0.0&4.18&1.65\\
2019 Nov 29 & C2PU & 2.01 & 1.99 & 27.93 & 353.1 & r & 16.88 & 0.03&1.88&2.40\\
2019 Dec 20 & ZTF & 2.03 & 1.94 & 28.63 & 8.2 & g & 17.50 & 0.06&2.51&1.90\\
\hline
\caption{Columns: (1) observation date; (2) observatory; (3) heliocentric distance; (4) topo-centric distance; (5) phase angle; (6) true anomaly; (7) filter; (8) 10$^4$ km aperture mag; (9) 1-$\sigma$ mag uncertainty, (10) in-image seeing of observations, (11) airmass of observations}
\end{longtable}

\end{CJK*}
\end{document}